\documentclass[twocolumn,twocolappendix]{aastex63}
\usepackage{ragged2e}
\usepackage{natbib}
\usepackage{scrextend}

\shorttitle{Molecular gas in type~1 AGNs}
\shortauthors{Molina, J. et al.}

\begin{document}

\title{Lack of Correlations between Cold Molecular Gas and AGN Properties in Type~1 AGNs at $z \lesssim 0.5$ }

\author[0000-0002-8136-8127]{Juan Molina}
\affil{Kavli Institute for Astronomy and Astrophysics, Peking University, Beijing 100871, China}
\affil{Department of Space, Earth and Environment, Chalmers University of Technology, Onsala Space Observatory, 439 92 Onsala, Sweden}

\author[0000-0002-4569-9009]{Jinyi Shangguan}
\affil{Max-Planck-Institut f\"{u}r Extraterrestrische Physik (MPE), Giessenbachstr., D-85748 Garching, Germany}

\author[0000-0003-4956-5742]{Ran Wang}
\affil{Kavli Institute for Astronomy and Astrophysics, Peking University, Beijing 100871, China}
\affiliation{Department of Astronomy, School of Physics, Peking University, Beijing 100871, China}

\author[0000-0001-6947-5846]{Luis C. Ho}
\affil{Kavli Institute for Astronomy and Astrophysics, Peking University, Beijing 100871, China}
\affiliation{Department of Astronomy, School of Physics, Peking University, Beijing 100871, China}

\author[0000-0002-8686-8737]{Franz E. Bauer}
\affil{Instituto de Astrof{\'{\i}}sica, Facultad de F{\'{i}}sica, Pontificia Universidad Cat{\'{o}}lica de Chile, Campus San Joaqu{\'{i}}n, Av. Vicu{\~{n}}a Mackenna 4860, Macul Santiago, Chile, 7820436} 
\affiliation{Centro de Astroingenier{\'{\i}}a, Facultad de F{\'{i}}sica, Pontificia Universidad Cat{\'{o}}lica de Chile, Campus San  Joaqu{\'{i}}n, Av. Vicu{\~{n}}a Mackenna 4860, Macul Santiago, Chile, 7820436}
\affiliation{Millennium Institute of Astrophysics, Nuncio Monse{\~{n}}or S{\'{o}}tero Sanz 100, Of 104, Providencia, Santiago, Chile} 

\author[0000-0001-7568-6412]{Ezequiel Treister}
\affil{Instituto de Astrof{\'{\i}}sica, Facultad de F{\'{i}}sica, Pontificia Universidad Cat{\'{o}}lica de Chile, Campus San Joaqu{\'{i}}n, Av. Vicu{\~{n}}a Mackenna 4860, Macul Santiago, Chile, 7820436} 
\affiliation{Centro de Astroingenier{\'{\i}}a, Facultad de F{\'{i}}sica, Pontificia Universidad Cat{\'{o}}lica de Chile, Campus San Joaqu{\'{i}}n, Av. Vicu{\~{n}}a Mackenna 4860, Macul Santiago, Chile, 7820436}

\begin{abstract} 
We present new NOrthern Extended Millimeter Array (NOEMA) observations of the CO(2--1) emission in eight of the brightest Palomar-Green quasars at $z \lesssim 0.5$ to investigate the role of active galactic nuclei (AGN) feedback in luminous quasars detected at low redshifts. We detect CO(2--1) emission in three objects, from which we derive CO luminosities, molecular gas masses and fractions, and gas depletion times. In combination with data available in the literature, we build a total sample of 138 local type~1 AGNs with CO(2--1) measurements.  We compare the AGN properties with the host galaxy molecular gas properties, considering the systems non-detected in CO emission.  We find that the CO luminosity does not correlate with AGN luminosity and Eddington ratio,  while the molecular gas fraction is weakly correlated with Eddington ratio.  The type~1 AGNs can be roughly separated into two populations in terms of infrared-to-CO luminosity ratio,  one population presenting values typically found in normal star-forming systems, while the other have lower ratio values, comparable to those measured for starbursts.  We find no evidence that AGN feedback rapidly quenches star formation in type~1 AGNs.  Our results may imply an underlying the role of host galaxy gravitational instabilities or the fast inflow of cold gas in triggering AGN activity.
\end{abstract}


\section{Introduction}

The evolution of galaxies is tightly correlated with the growth of the supermassive black holes (BHs) as evidenced by the correlations between the BH mass and the host galaxy bulge properties \citep{Magorrian1998,Ferrarese2000,Gebhardt2000}. The connection between BH accretion and the host galaxy growth naturally arises because both depend on the same fuel reservoir, replenished largely by similar processes that drive gas inward \citep{Kormendy2004}. The co-evolution of the BH and host galaxy \citep{Kormendy2013} is widely accepted to be regulated by the feedback of the active galactic nuclei (AGN; \citealt{Fabian2012}). During the accretion of mass onto the BH, vast amounts of energy are released, and a merely minor fraction of this power coupled to the BH surroundings can heat and/or expel the gas from the host galaxy (e.g., \citealt{Somerville2008,Schaye2015,Sijacki2015,Nelson2018}), thereby halting the ongoing star formation activity \citep{Dubois2016}. The AGN feedback can also affect the galaxy halo, preventing the condensation of cold and warm gas, and further hamper the star formation over long timescales \citep{Bower2006,Croton2006,Fabian2012,Gaspari2020}. 

During the last decade, numerous efforts have been focused to test whether AGN feedback effectively removes sufficient cold gas from the host galaxy to curtail ongoing star formation activity. Studies focusing on comparisons with control samples and/or the main-sequence of star-forming galaxies have reported no difference between the inactive galaxies and those hosting AGNs (\citealt{Rosario2018,Kirkpatrick2019,Schulze2019,Florez2020,Koss2021,Salvestrini2022,Smirnova-Pinchukova2022}, but see \citealt{Vito2014, Scholtz2018,Stemo2020}). Correlations between key host galaxy properties, such as molecular hydrogen (H$_2$) content, star formation rate (SFR), and/or star formation efficiency, with AGN luminosity suggest that low-$z$ luminous AGNs are mainly hosted in typical star-forming galaxies (e.g., \citealt{Harrison2012,Rosario2012,Rosario2013,Zhang2016,Stanley2017,Husemann2017,Bernhard2019,Grimmett2020,Yesuf2020,Jarvis2020,Koss2021,Vietri2022}), while the more luminous active systems tend to be found in starbursts (e.g., \citealt{Young2014,Bernhard2016,Pitchford2016,Kirkpatrick2020,Shangguan2020b,Xie2021,Zhuang2021}). Only a few studies of high-$z$ AGNs ($z \approx 1-2$) suggest reduced molecular gas content in AGN hosts \citep{Kakkad2017,Perna2018,Circosta2021,Bischetti2021}. Whether or how the luminous AGN does efficiently remove or heat the gas from host galaxy and affect the ongoing star formation is still uncertain \citep{Harrison2017}. 

\begin{figure}
\centering
\includegraphics[width=1.0\columnwidth]{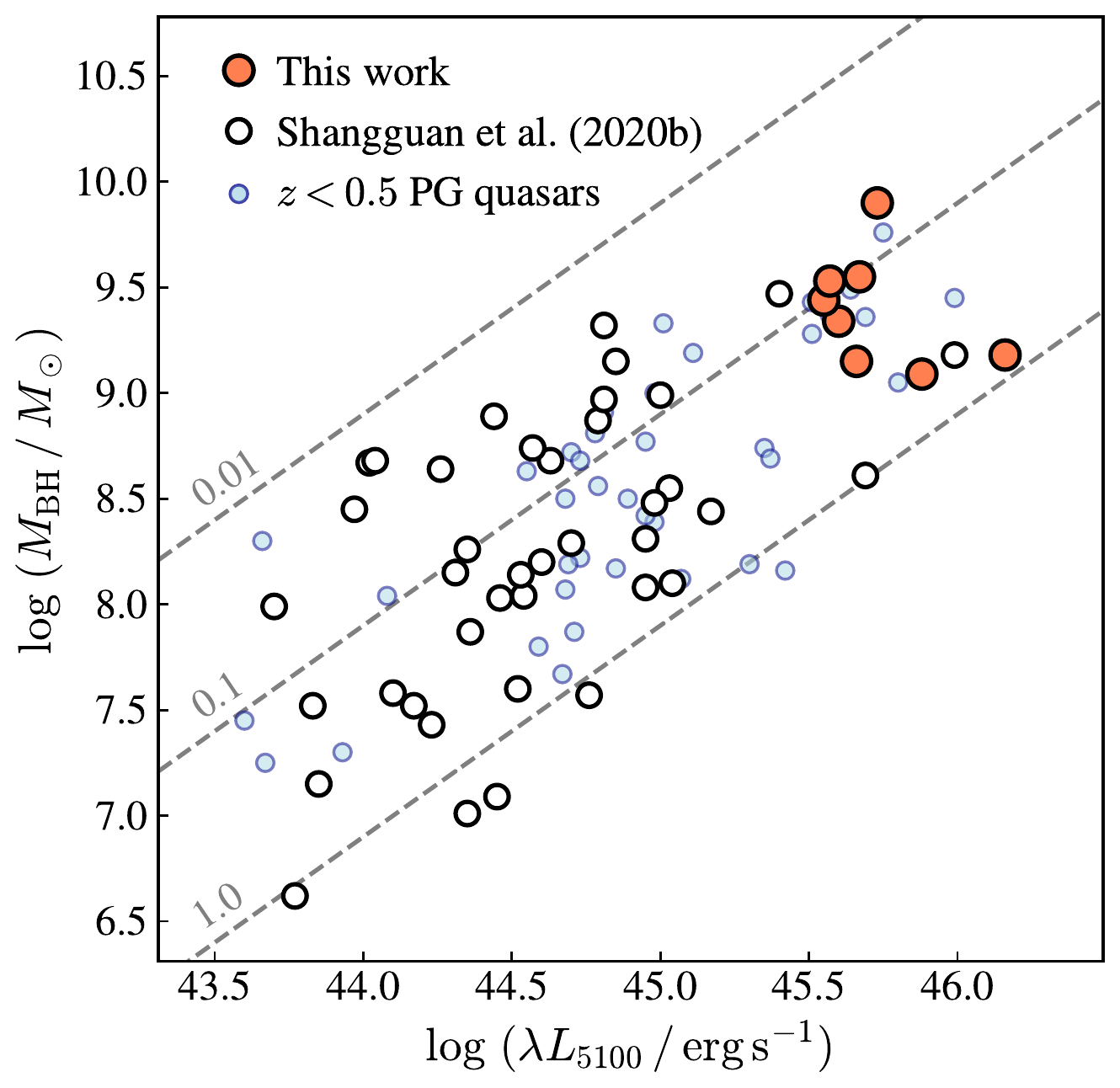}
\caption{\label{fig:NOEMA_sample} Black hole mass as a function of the AGN monochromatic luminosity at 5100\,\r{A} for the PG quasar host galaxies. The red-filled circles present the 8 targets observed by NOEMA and analyzed in this work. The open circles show the PG quasars with previous CO observations \citep{Shangguan2020}, while the blue circles correspond to the parent sample of $z \lesssim 0.5$ PG quasars observed by Herschel \citep{Shangguan2018}.  The dashed lines represent constant Eddington ratio values.  We target a sub-sample of PG quasars with more massive BHs and luminous AGNs, complementing the previous CO observations.}
\end{figure}

Optically visible and largely unobscured quasars--the most luminous AGNs-- allows us to study possible effects of AGN feedback on their host galaxies. An unobscured quasar is thought to arise after the overwhelming release of energy from the AGN that expels the gas and dust enshrouding the nucleus (e.g., \citealt{Sanders1988,Hopkins2008}).  Strikingly,  the study of the global IR ($\sim 1 - 500\,\mu$m) spectral energy distribution (SED) emission in a set of 87 local ($z < 0.5$) Palomar-Green (PG) quasars \citep{Boroson1992} suggested that most of these unobscured AGNs have dust-based molecular gas masses comparable to normal star-forming galaxies with similar stellar mass \citep{Shangguan2018}. Follow-up Atacama compact array (ACA) observations targeting the carbon-monoxide molecule ($^{12}$CO) emission and previous literature observations confirmed the results from the global IR SED analysis.  The CO$\,J = 2 \rightarrow 1$ transition [$\nu_{\rm rest} = 230.538\,$GHz; hereafter CO(2--1)] emission was detected in a total 34 out of 40 PG quasars. The global CO emission line shapes further suggested that the molecular gas appears kinematically regular within the host galaxies (\citealt{Shangguan2020b,Shangguan2020}, see also \citealt{Molina2021}). However, the PG quasar sample observed in CO is admittedly biased toward the less luminous local PG quasars ($L_{\rm bol} \lesssim 10^{45}\,$erg\,s$^{-1}$; Figure~\ref{fig:NOEMA_sample}). Luminous AGNs can drive multi-phase outflows and deposit its energy into the host galaxy interstellar medium (ISM; \citealt{Cicone2014,Feruglio2015,Perna2015,Morganti2017,Cicone2018,Fluetsch2019,Veilleux2020,Girdhar2022}). The outflow energetics correlate with AGN luminosity (molecular gas mass outflow rates $\propto L_{\rm bol}^{0.76}$ and kinetic power $\propto L_{\rm bol}^{2.27}$; \citealt{Fiore2017}) meaning that the AGN feedback effect on the host galaxy ISM, if any, should be more easily detected in more luminous AGNs.

In this work, we report new NOrthern Extended Millimeter Array (NOEMA) observations targeting the CO(2--1) for a sub-sample of eight $z \approx 0.3-0.5$ PG quasars selected from \citet{Shangguan2018}.  Those targets correspond to the more luminous PG quasars at $z \lesssim 0.5$.  By using complementary data taken from the literature,  we build a total sample of 138 unobscured local AGNs covering the $L_{\rm bol} \approx  10^{43}-10^{47}\,$erg\,s$^{-1}$ range to study the relation between the host galaxy molecular gas content and AGN activity, and to explore whether AGN feedback is effectively operating in those sources.  The paper is organized as follows. Section~\ref{sec:obs} presents our sample, observation strategy and data reduction. The data analysis is described in Section~\ref{sec:met}, and the subsequent results are described in Section~\ref{sec:res}. We further discuss our findings in Section~\ref{sec:dis}. Section~\ref{sec:con} is the conclusion. We adopt a $\Lambda$CDM cosmology with $\Omega_m = 0.308$, $\Omega_\Lambda = 0.692$, and $H_0 = 67.8\,$km\,s$^{-1}\,$Mpc$^{-1}$ \citep{Planck2016}.

\begin{table*}
	\centering
	\def\arraystretch{1.2}
	\setlength\tabcolsep{4pt}
    	\caption{\label{tab:sample} Basic Parameters of the Sample}
    	\vspace{0.2mm}
    	\hspace{-2.5cm}
	\begin{tabular}{ccccccccc}
		\hline
		\hline
		Object & R.A. & Decl. & $z$ & Morphology & $\log M_\star$ & $\log \lambda L_{\rm 5100}$ & $\log M_{\rm BH}$ & $L_{\rm bol}/L_{\rm Edd}$ \\
		& (J2000.0) & (J2000.0) & & & ($M_\odot$) & (erg\,s$^{-1}$) & ($M_\odot$) & \\
		(1) & (2) & (3) & (4) &(5) & (6) & (7) & (8)& (9) \\
		\hline
			PG\,1049$-$005 & 10:51:51.45 & $-$00:51:17.7 & 0.357 & Merger & 11.1$^*$ & 45.60 & 9.34 & 0.14 \\ 
			PG\,1100$+$772 & 11:04:13.86 & $+$76:58:58.2 & 0.313 & Elliptical & 11.27 & 45.55 & 9.44 & 0.10 \\
			PG\,1259$+$593 & 13:01:12.96 & $+$59:02:06.7 & 0.472 & Elliptical & 10.99 & 45.88 & 9.09 & 0.49 \\
			PG\,1425$+$267 & 14:19:03.82 & $+$26:32:14.5 & 0.366 & Elliptical & 11.15 & 45.73 & 9.90 & 0.05 \\
			PG\,1512$+$370 & 15:14:43.07 & $+$36:50:50.3 & 0.371 & Elliptical & 11.01 & 45.57 & 9.53 & 0.09 \\
			PG\,1704$+$608 & 17:04:41.37 & $+$60:44:30.5 & 0.371 & Elliptical & 11.52 & 45.67 & 9.55 & 0.10 \\
			PG\,2112$+$059 & 21:14:52.58 & $+$06:07:42.4 & 0.466 & Elliptical & 10.8$^*$ & 46.16 & 9.18 & 0.76 \\ 
			PG\,2251$+$113 & 22:54:10.42 & $+$11:36:38.7 & 0.323 & Elliptical & 11.05 & 45.66 & 9.15 & 0.26 \\
		\hline
	\end{tabular}
	\justify
	{\textsc{Note}--- (1) Source name. (2) Right ascension. (3) Declination. (4) Redshift. (5) Morphology of the host galaxy \citep{Zhang2016,Zhao2021} (6) Stellar mass; the $1\,\sigma$ uncertainty is 0.3\,dex \citep{Shangguan2018}.  (7) AGN monochromatic luminosity at 5100\,\r{A}. We assume a $1\sigma$ error of 0.05\,dex \citep{Vestergaard2006}. (8) Black hole mass, estimated by applying the calibration of \citet{Ho2015} and taken from \citet{Shangguan2018}; the $1\sigma$ uncertainty is 0.3\,dex. (9) Eddington ratio, where $L_{\rm bol} = 10\, L_{\rm 5100}$ \citep{Richards2006} is the bolometric luminosity, and $L_{\rm Edd} = 1.26 \times 10^{38} (M_{\rm BH}/M_\odot)$\,erg\,s$^{-1}$ is the Eddington luminosity. ($^*$) The stellar mass values are taken from \citet{Zhao2021}, but carry large uncertainty ($\sim 0.7-0.9\,$dex).}
\end{table*} 

\begin{table*}
 	   \tiny
        \centering
        \def\arraystretch{1.2}
        \setlength\tabcolsep{3pt}
        \caption{\label{tab:obssetup} NOEMA Observational Setup}
        \vspace{0.2mm}
        \begin{tabular}{crllcccccc}
                \hline
                \hline
                Object & Observation\,\,\,\, & \,\,\,\,\,\,\,\,\,\,\,\,\,\,Bandpass \& & \,\,\,\,\,\,\,\,\,\,\,\,\,\,\,\,\,\,\,\,Amplitude \& & PWV & Observing & Beam Size & Beam Position & Half-power & Maximum Recoverable  \\
                 &  Date\,\,\,\,\,\,\,\,\,\,\,\,\, & \,\,\,\,\,\,\,\,\,Flux Calibrator & \,\,\,\,\,\,\,\,\,\,\,\,\,\,\,Phase Calibrator & (mm) & Time$^*$ (h) & ($\arcsec \times \arcsec $) & Angle ($^\circ$) & Beamwidth ($\arcsec$) & Scale$^{**}$ ($\arcsec$) \\
                (1) & (2)\,\,\,\,\,\,\,\,\,\,\,\,\,\,\, &   \,\,\,\,\,\,\,\,\,\,\,\,\,\,\,\,\,\,\,\,\,\,\,\,\,(3) &  \,\,\,\,\,\,\,\,\,\,\,\,\,\,\,\,\,\,\,\,\,\,\,\,\,\,\,\,\,\,\,\,\,\,(4) & (5) & (6) & (7) & (8) & (9) & (10) \\
                \hline
		PG\,1049$-$005 & 10\,Apr.\,2022 & 3C84 & 1044+018\,/\,J1028$-$0236 & 3 & 1.5 & $3.3 \times 1.0$ & 11 & 29.7 & 9.9 \\
        		PG\,1100$+$772 & 6/14\,Apr.\,2022 & 3C345 & 1039+811\,/\,1044+719 & 4 & 4.1 & $2.8 \times 1.1$ & 40 & 28.7 & 12.1 \\
        		PG\,1259$+$593 & 13\,Apr.\,2022 & 3C279 & 1300+580\,/\,J1259+516 & 5--6 & 2.6 & $2.4 \times 1.5$ & 127 & 32.2 & 12.8 \\
        		PG\,1425$+$267 & 6\,Apr.\,2022 & 3C273 & 1417+273\,/\,J1422+323 & 4 & 1.1  & $2.4 \times 1.3$ & 110 & 29.9 & 14.5 \\ 
        		PG\,1512$+$370 & 10\,Apr.\,2022 & 3C279 & 3C345\,/\,1504+377\,/\,1505+428 & 2 & 2.6 & $2.4 \times 1.0$ & 71 & 30.0 & 11.3 \\
        		PG\,1704$+$608 & 14/15\,Apr.\,2022 & 0851+202\,/\,2013+370 & 1637+574\,/\,1645+635 & 3--6 & 2.2 & $3.0 \times 1.3$ & 98 & 30.0 & 11.7 \\
        		PG\,2112$+$059 & 6\,Apr.\,2022 & 3C345 & 2059+034\,/\,2121+053 & 4 & 1.5 & $3.5 \times 1.4$ & 18 & 32.0 & 13.8 \\
        		PG\,2251$+$113 & 17/18\,Apr.\,2022 & 2013+270\,/\,2230+114  & 3C454.3\,/\,2230+114 & 1--3 & 6.4 & $3.7 \times 1.4$ & 12 & 29. 5 & 14.5 \\      
                \hline
        \end{tabular}
        \justify
	{\small \textsc{Note}--- (*) Including overheads. }
\end{table*}

\section{Sample and Observations}
\label{sec:obs}

We focus on observing 8 PG quasars taken from the broader sample of 87 $z < 0.5$ quasars belonging to the PG survey \citep{Boroson1992}. The PG quasars were selected by their optical/ultraviolet colors, and correspond to luminous, broad-lined (type~1) AGNs. Multi-wavelength data are available across the entire electromagnetic spectrum range, from X-ray \citep{Reeves2000,Bianchi2009} through optical \citep{Boroson1992,Ho2009}, mid-IR \citep{Shi2014,Xie2021,Xie2022}, far-IR \citep{Petric2015,Shangguan2018,Zhuang2018}, mm \citep{Shangguan2020b,Shangguan2020}, and radio \citep{Kellermann1989,Kellermann1994,Silpa2020} wavelengths, allowing detailed SED characterization and accurate estimation of the global SFR and gas content of the host galaxies \citep{Shangguan2018}. A major fraction of the sample has been already observed in the optical and near-IR by the Hubble Space Telescope (HST) at $\sim 0.1\arcsec$ resolution \citep{Kim2008,Zhang2016,Kim2019,Zhao2021}, while arcsecond resolution interferometric and integral field unit data are accessible for a few of these \citep{Molina2021,Molina2022}. 

\citet{Shangguan2020b} presented CO(2--1) observations for a representative sample of 40 $z<0.3$ PG quasars,  23 newly observed host galaxies using ACA (synthesized beam FWHM $\sim 6\, \arcsec$).  They reported line emission detection in 34 host galaxies, 21 sources detected from their ACA campaign.  For the ACA data,  \citet{Shangguan2020b} computed the global CO velocity integrated fluxes by spatially collapsing the spectra over the $2\,\sigma$ contour region of the imaged data, to then fit the emission line using a single Gaussian or double-horn profile functions.  

We selected the 8 PG quasar host galaxies to complement and expand the sample already observed in CO(2--1) emission, specially at high AGN luminosity ($\lambda L_{5100} \gtrsim 10^{45.5}\,$\,erg\,s$^{-1}$) where \citet{Shangguan2020b} only reported CO(2--1) measurements for 3 host galaxies (Figure~\ref{fig:NOEMA_sample}). Our selection was based on requiring: (1) PG quasars with AGN monochromatic luminosities at 5100\,{\AA} ($\lambda L_{5100}$) higher than $10^{45.5}\,$\,erg\,s$^{-1}$; and (2) targets with robust AGN/host galaxy far-IR SED decomposition performed by \citet{Shangguan2018}. The first criterion implies that we build a combined sample of PG quasars spanning $\sim 2$ orders of magnitude in terms of $\lambda L_{5100}$ and BH masses (Figure~\ref{fig:NOEMA_sample}).  The second criterion ensures that the host galaxies have accurate estimates for the dust 8 to 1000$\mu$m far-IR luminosity,  so that their SFR could be reliably computed.  We show the properties of our targets in Table~\ref{tab:obssetup}.

The CO(2--1) emission line was observed with NOEMA on 6--18 April, 2022, as part of the program W21CI (PI:~J. Molina). The observations were carried out using 10 or 11 antennas in compact array (D) configuration,  dual polarization, and for total sample observing time of 22\,h. We used the autocorrelator PolyFix, which covers a total bandwidth of 15.5 GHz in each linear polarization, and split between the lower sideband (LSB) and the upper sideband (USB). The channel width is 2 MHz, namely 3.4\,km\,s$^{-1}$ at the tuning frequency (set in the USB) of our observations.  The synthesized beam FWHM size is $1.0-3.7\, \arcsec$, corresponding to scales between 5 and 21\,kpc at the redshift of our sources.  The observational setup for each source is detailed in Table~\ref{tab:obssetup}.

\subsection{Data Reduction}
\label{sec:data_red}

We reduced the data following the standard procedure using the GILDAS software \citep{Guilloteau2000}. The data were calibrated using the Institute for Radio Astronomy in the Millimeter Range (IRAM) package Continuum and Line Interferometer Calibration (CLIC). The standard pipeline reduction and calibration was implemented to a large extent, but some minor fraction of data scans were flagged following the astronomer on duty report.\footnote{https://www.iram.fr/IRAMFR/GILDAS/doc/pdf/pdbi-cookbook.pdf\label{fn:cookbook}} We imaged the data using the MAPPING software of GILDAS. We used the \texttt{uv\_cont} task to combine all the line-free channels in both sidebands (we also avoided the noisy channels), and then we merged those using \texttt{uv\_merge} to produce the continuum $uv$ table for each source. To produce the emission line $uv$ table we subtracted the continuum using \texttt{uv\_base} from the calibrated $uv$ tables in the USB.  Then,  we used the task \texttt{uv\_compress} to set the data spectral resolution.  For PG\,1049$-$005 and PG\,1259$+$593 we set the channel width $\sim 40\,$km\,s$^{-1}$ as both were detected in CO emission at high signal-to-noise.  For the rest of the sample we set a channel width $\sim 100\,$km\,s$^{-1}$ after exploring different spectral setup aiming to maximize CO line detection.\footnote{The target PG\,2251+113 was found offset by $2.2\arcsec$ and $-2.7\arcsec$ in right ascension and declination, respectively, from the expected location. We corrected the observation pointing by this offset using \texttt{uv\_shift} before proceeding with the continuum and emission line $uv$ tables construction.} For both, the continuum and the emission line $uv$ tables,  we produced the dirty images using the task \texttt{uvmap} and estimated the data rms level (Table~\ref{tab:CO_prop}). Those values broadly correspond to our observation sensitivity requirements.  The image pixel size is set to $0.2\arcsec$ following \texttt{uvmap} task recommendation.  We used the \texttt{CLEAN} algorithm in the version of \citet{Hogbom1974} to do the data deconvolution. We use natural weighting to maximize the observation sensitivity.  We set \texttt{CLEAN} down to the data rms level by doing many iterations as needed.  After doing the cleaning, we visually checked the residual maps, finding no emission higher than the noise level in any target.  The data are corrected for primary beam.  Finally, the imaged data was converted to the usual \texttt{FITS} format using the task \texttt{vector$\backslash$fits}.  We consider a conservative systematic NOEMA flux calibration uncertainty of 20\,\% at 2\,mm.\footref{fn:cookbook}

\subsection{IR Spectral Energy Distribution}
\label{sec:farIR_SED}

To estimate the IR luminosity decontaminated from AGN effects for our targets,  we benefit from the panchromatic IR SED from $\sim 1$ to 500\,$\mu$m, complemented by mid-IR ($5-38\,\mu$m) spectra from the Spitzer Infrared Spectrometer (IRS), and available radio data collated in \citet{Shangguan2018}.  Briefly,  in the near-IR the fluxes were re-derived by \citet{Shangguan2018} from the 2MASS \citep{Skrutskie2006} $J$,  $H$ and $K_s$ bands \citep{Cohen2003} images using $7\,\arcsec$ aperture radius with the sky annulus set to a radius of 25 to $35\,\arcsec$.  \citet{Shangguan2018} also re-estimated the mid-IR WISE \citep{Wright2010,Jarrett2011} $W1$ ($3.353\,\mu$m) and $W2$ ($4.603\,\mu$m) fluxes for the PG quasars, but avoided to include the $W3$ ($11.561\,\mu$m) and $W4$ ($22.088\,\mu$m) data due to known systematic uncertainties associated to the red color of the targets, plus the bands wavelength overlap with the bandpass of the Spitzer IRS spectra.  The WISE $W1$ and $W2$ fluxes were computed using a similar method to that used for the 2MASS images.  An aperture radius of $8.25\,\arcsec$, plus a sky annulus of $50-70\,\arcsec$ were adopted.  Coadded PSFs for the WISE bands were used to compute the aperture correction factors.  The Herschel Photodetector Array Camera and Spectrometer (PACS, \citealt{Poglitsch2010}) and Spectral and Photometric Imaging Receiver (SPIRE; \citealt{Griffin2010}) observations were formally presented in \citet{Shangguan2018} for the PG quasars.  For the PACS data,  point-source aperture photometry were performed using 12,  12, and $22\,\arcsec$ aperture radius for the 70, 100, and $160\,\mu$m bands.  Sky annulus radius were set to $35-45\,\arcsec$ to minimize PSF wing contamination,  and aperture corrections were also applied.  From the SPIRE data,  the sources flux density values or upper limits were computed following \citet{Savage2007}.  The Spitzer data were processed by \citet{Shi2014},  who scaled the short-low ($5-14\,\mu$m) spectrum,  the long-low ($14-40\,\mu$m) spectrum, and the overall flux of the spectrum to match the Spitzer MIPS $24\,\mu$m photometry.  \citet{Shangguan2018} also found that the density flux scale of the Spitzer spectra were well-matched with respect to the WISE data.  The radio data were updated from that presented in \citet{Shangguan2018} considering newer literature data.  We present the radio data in Appendix~\ref{sec:AppA}. 

\begin{figure*}
\centering
\includegraphics[width=2.0\columnwidth]{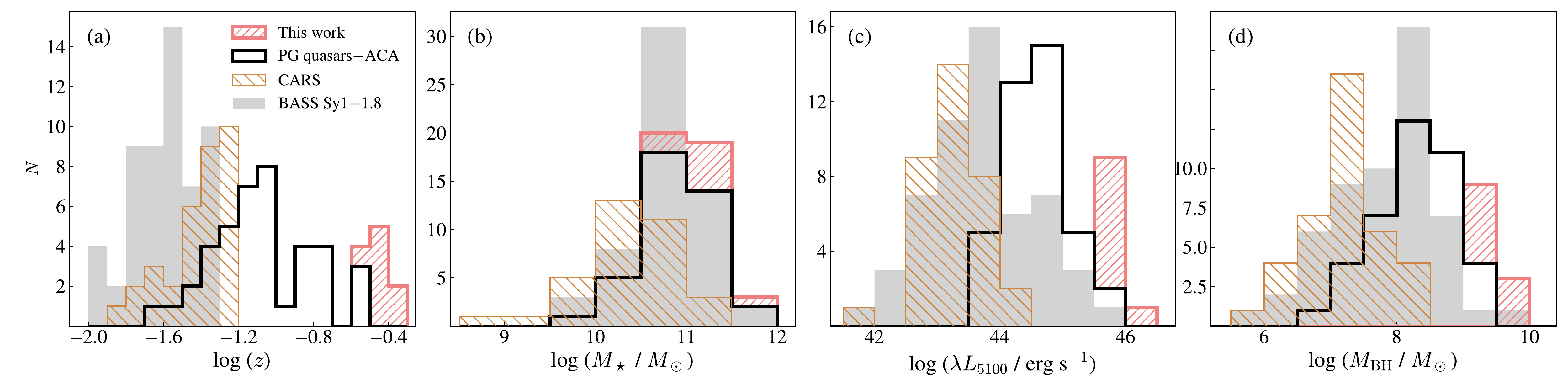}
\caption{\label{fig:comp_sample} Comparison between our target sample (red) with the sample of PG quasars with previously detected CO emission (black; \citealt{Shangguan2020}), the BASS type~1 AGN sub-sample (grey),  and the CARS survey (brown),  in terms of the (a) redshift, (b) host galaxy stellar mass,  (c) 5100\,\r{A} AGN luminosity,  and (d) BH mass.  Our target sample is presented over the ``PG quasar--ACA'' sample data to improve figure visualization.}
\end{figure*}

\subsection{Complementary AGN Sample}
\label{sec:comp_sample}

We complement our PG quasar CO survey with literature data taken from two main surveys,  the BAT-AGN Spectroscopic Survey (BASS) sample presented in \citet{Koss2022} and the Close AGN Reference Survey (CARS; \citealt{Husemann2022}).  In their data-release 2,  the BASS AGNs consist of 858 nearby ($z \lesssim 0.24$) AGNs selected by their ultrahard (14--195\,keV) X-ray emission.  We select a sub-sample of BASS AGNs presenting H$\alpha$ and H$\beta$ broad-lines,\footnote{\citet{MejiaRestrepo2022} refer to those systems as Seyfert~1,  1.2, 1.5, and 1.8, based on the ratio between the broad H$\beta$ and [O\,{\sc iii}] line luminosity. Here, we only refer to those BASS AGNs as ``Sy1--1.8''.} but we discard all the BASS AGNs with poor optical spectra models following \citet{MejiaRestrepo2022}.  From this sub-sample, we further select the host galaxies with CO(2--1) line emission detection or upper limit.  Those BASS AGN host galaxies were observed with the Atacama Pathfinder Experiment telescope (APEX; 46 sources), the James Clerk Maxwell Telescope (JCMT; 7 sources) and the Caltech Submillimeter Observatory (CSO; 3 sources).  At the observed frequencies ($\sim 220-229\,$GHz), the APEX, JCMT, and CSO observations have effective primary beam size of $26.3-27.5\,\arcsec$, $20.4\,\arcsec$,  and $32\,\arcsec$, respectively,  corresponding to scales of $\sim 6-27\,$kpc at the sources redshift \citep{Koss2021}.  Specifically, we adopt the CO velocity integrated fluxes computed by fitting the line profiles with a single Gaussian or double-horn functions.  We note that \citet{Koss2021} show that those values are in good agreement with estimates computed by simply summing the CO emission over the line spectral range (see their Appendix D).  The BASS AGN measurement uncertainties are 10\,\% and 15\,\% for flux calibration and aperture correction, respectively.  We take the BASS AGN host galaxy properties from \citet{Koss2021}. For the CARS survey, we take the AGN and host galaxy properties from \citet{Husemann2022} and \citet{Smirnova-Pinchukova2022},  respectively.  We note that the CO(2--1) line flux measurements were carried out using the IRAM 30--m telescope \citep{Bertram2007}.  The IRAM 30--m telescope primary beam size is $\sim 10.9-11.4\,\arcsec$ at the observed frequencies ($\sim217-227\,$GHz), and correspond to a physical scale $\sim 3-11\,$kpc.  \citet{Bertram2007} estimate the CO line velocity integrated fluxes by simply summing the flux density over the full velocity range of the emission line. The IRAM 30--m telescope flux calibration uncertainty is $\sim10$\,\%.

In consistency with the estimates presented for the PG quasars (Table~\ref{tab:sample}),  for both BASS and CARS AGN surveys we use the AGN monochromatic luminosity at 5100\,\r{A} and broad H$\beta$ line width values to estimate $L_{\rm bol}$ and $M_{\rm BH}$ following \citet{Richards2006} and \citet{Shangguan2018}, respectively.  The host galaxy CO luminosity and SFR values are estimated by using the conventions adopted in this work, and the stellar masses are scaled assuming the \citet{Kroupa2001} initial mass function (IMF; see Section~\ref{sec:met}).  The BH masses estimated by us and the values reported by \citet{MejiaRestrepo2022} for the BASS AGNs are in good agreement with an average BH mass ratio of $\sim 0.9$ and 0.25\,dex scatter. For the CARS sample there are several systems with fewer than two SED data points at $\lambda \gtrsim 60\mu$m,  implying inaccurate far-IR emission luminosity estimates.  Those correspond to 19 out of 38 host galaxies that were excluded whenever $L_{\rm IR}$ was involved in our analysis.  On average, the discarded systems have slightly lower stellar mass ($-0.2$\,dex) and CO(2--1) luminosity ($-0.1$\,dex) compared to the full CARS sample,  but present no systematic offset in terms of average molecular gas fraction,  bolometric luminosity,  BH mass, and Eddington ratio estimates (offset $\lesssim 0.05$\,dex in terms of absolute value).  However,  we caution that by discarding those CARS host galaxies we may result with a sub-sample biased toward the most IR luminous systems.  The excluded CARS AGNs tend to be found at slightly higher redshifts ($z \approx 0.05-0.06$) compared to the sample average ($\langle z \rangle \approx 0.04$).  We further exclude few duplicated objects among the BASS, CARS and PG quasar surveys. 

Figure~\ref{fig:comp_sample} presents the properties of our sample, the PG quasars with previous CO line measurements \citep{Shangguan2020},  the sub-sample of broad-lined BASS AGNs, and the CARS targets.  The BASS sub-sample correspond 56 $z \lesssim 0.05$ host galaxies (21 upper-limit estimates) . The CARS sub-sample consists of 34 $z \lesssim 0.06$ systems (9 upper-limit estimates).  The PG quasars and BASS AGNs cover similar host galaxy stellar mass range,  while the CARS host galaxies tend to be less massive.  Compared to those three AGN samples, our targets tend to be slightly biased toward the more massive systems in terms of stellar mass, expected as we selected the local quasars with the most massive BHs, hence the host galaxies presenting the most massive bulges.  The sample of 8 PG quasars presented here probes the high BH mass and AGN luminosity tail of the $z \lesssim 0.5$ AGN population.

\begin{table*}
	\centering
	\def\arraystretch{1.2}
	\setlength\tabcolsep{4pt}
    	\caption{\label{tab:CO_prop} Flux Density and Derived Properties of the Sample.}
    	\vspace{0.2mm}
	\begin{tabular}{cccccccccccc}
		\hline
		\hline
		Object & $S_{\rm cont}$ & $\sigma_{\rm cont}$ & $S_{\rm CO} \Delta v$ & $\nu_{\rm obs}$ & FWHM & $\sigma_{\rm CO}$ & $\log L_{\rm IR}$ & $\log L'_{\rm CO(2-1)}$ & $\log M_{\rm H_2}$ \\
		 & (mJy) & (mJy\,beam$^{-1}$) & (Jy\,km\,s$^{-1}$) &  GHz & km\,s$^{-1}$ & (mJy) & (erg\,s$^{-1}$) & (K\,km\,s$^{-1}$\,pc$^2$) & $M_\odot$ \\ 
		(1) & (2) & (3) & (4) &(5) & (6) & (7) & (8) & (9) & (10) \\
		\hline
		PG\,1049$-$005 & $0.52\pm 0.11$ & 0.04 & $2.53 \pm 0.51$ & 170.300 & $181 \pm 5$ & 0.52 & 45.4 & 9.64$\pm$0.09 & 10.13 \\
        		PG\,1100$+$772 & $22.37\pm 4.47$ & 0.24 & $\lesssim 2.09$ & \nodata & \nodata & 2.00 & 44.8 & \nodata & \nodata \\
        		PG\,1259$+$593 & $\lesssim 1.78$ & 0.03 & $0.73\pm 0.16$ & 157.403 & $111 \pm 10$ & 0.75 & 44.7 & 9.35$\pm$0.10 & 9.85 \\
        		PG\,1425$+$267 & $4.69\pm 0.94$ & 0.06 & $\lesssim 0.59$ & \nodata & \nodata & 0.54 & 44.9 & \nodata & \nodata \\
        		PG\,1512$+$370 & $5.75\pm 1.15$ & 0.04 &  $\lesssim 0.39$ & \nodata & \nodata & 0.36 & 44.6 & \nodata & \nodata \\
        		PG\,1704$+$608 & $3.34\pm 0.67$ & 0.07 &  $\lesssim 0.83$ & \nodata & \nodata & 0.77 & 45.1 & \nodata & \nodata \\
        		PG\,2112$+$059 &  $\lesssim 2.40$ & 0.05 & $0.75 \pm 0.33$ & 157.093 & $398 \pm 169$ & 0.68 & 45.2 & 9.35$\pm$0.17 & 9.85 \\ 
        		PG\,2251$+$113 & $8.98\pm 1.80$ & 0.13 & $\lesssim 0.52$ & \nodata & \nodata & 0.49 & 44.6 & \nodata & \nodata \\
		\hline
	\end{tabular}
	\justify
	{\textsc{Note}--- (1) Source name. (2) Millimeter continuum flux density at restframe 230\,GHz. (3) Noise level of the continuum image. (4) CO(2--1) velocity integrated intensity. (5) CO(2--1) emission line observed frequency. (6) Full width at half maximum. (7) Noise level of the cleaned data cube. (8) Total infrared luminosity of the host galaxy calculated following \citet{Shangguan2018}. The typical uncertainty is $\sim 0.2-0.3$\,dex.  (9) CO(2--1) line luminosity. (10) Molecular gas mass derived from the CO luminosity, assuming $\alpha_{\rm CO} = 3.1\, M_\odot \,($K\,km\,s$^{-1}$\,pc$^2)^{-1}$. The 1$\sigma$ errors are dominated by the CO-to-$H_2$ conversion factor uncertainty ($\sim$0.3\,dex).  We note that the error estimates of the mm continuum flux density,  the CO(2--1) line velocity integrated intensity, and line luminosity consider a conservative NOEMA flux calibration uncertainty of $20$\,\% at 2\,mm.}
\end{table*}

\section{Methods}
\label{sec:met}

\subsection{CO(2--1) Emission-line and Millimeter Continuum Characterization}
\label{sec:data_charac}

\begin{figure*}
\centering
\includegraphics[width=0.49\columnwidth]{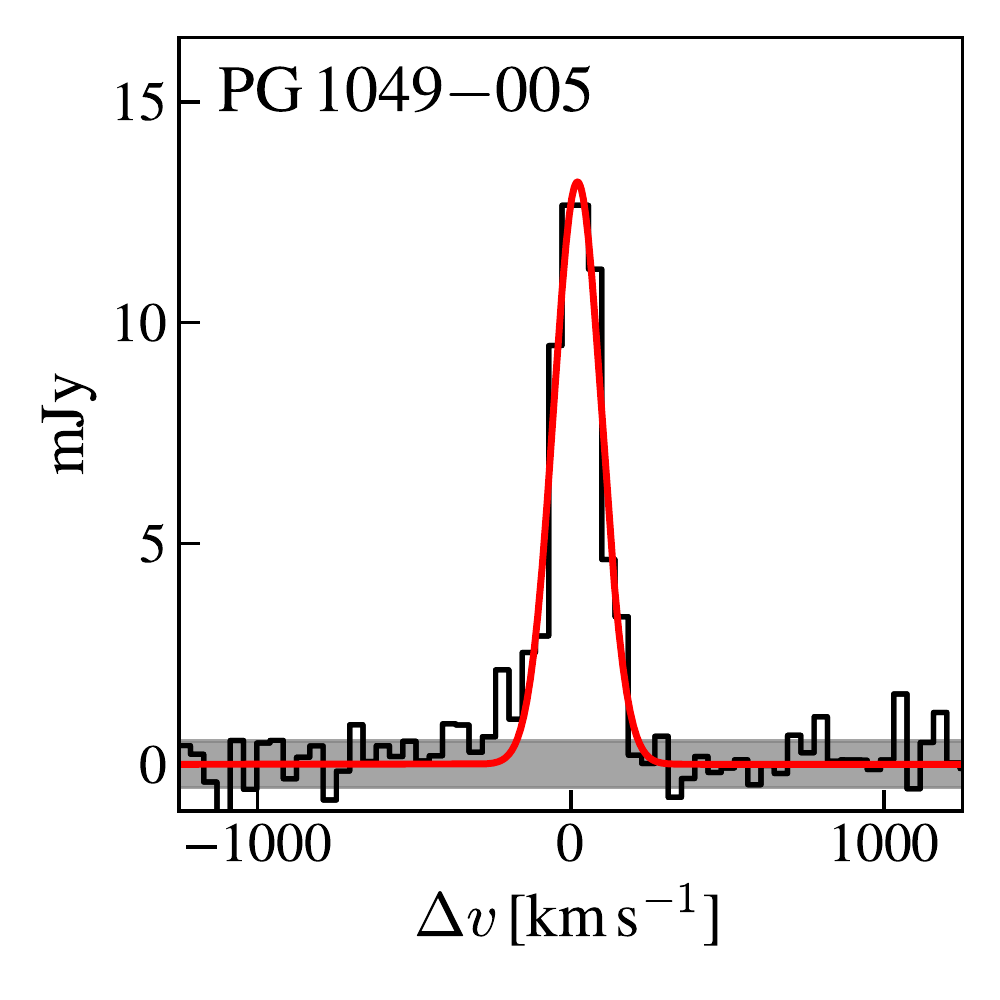}
\includegraphics[width=0.49\columnwidth]{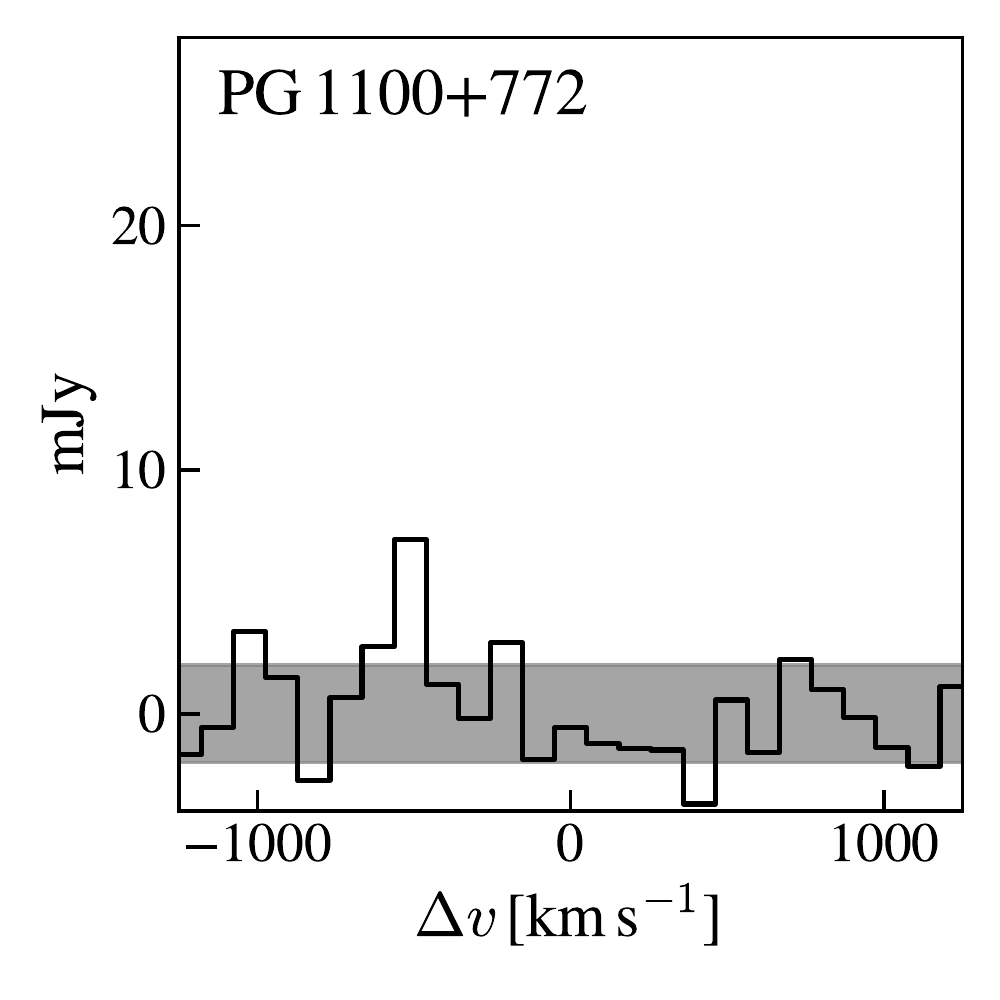}  
\includegraphics[width=0.49\columnwidth]{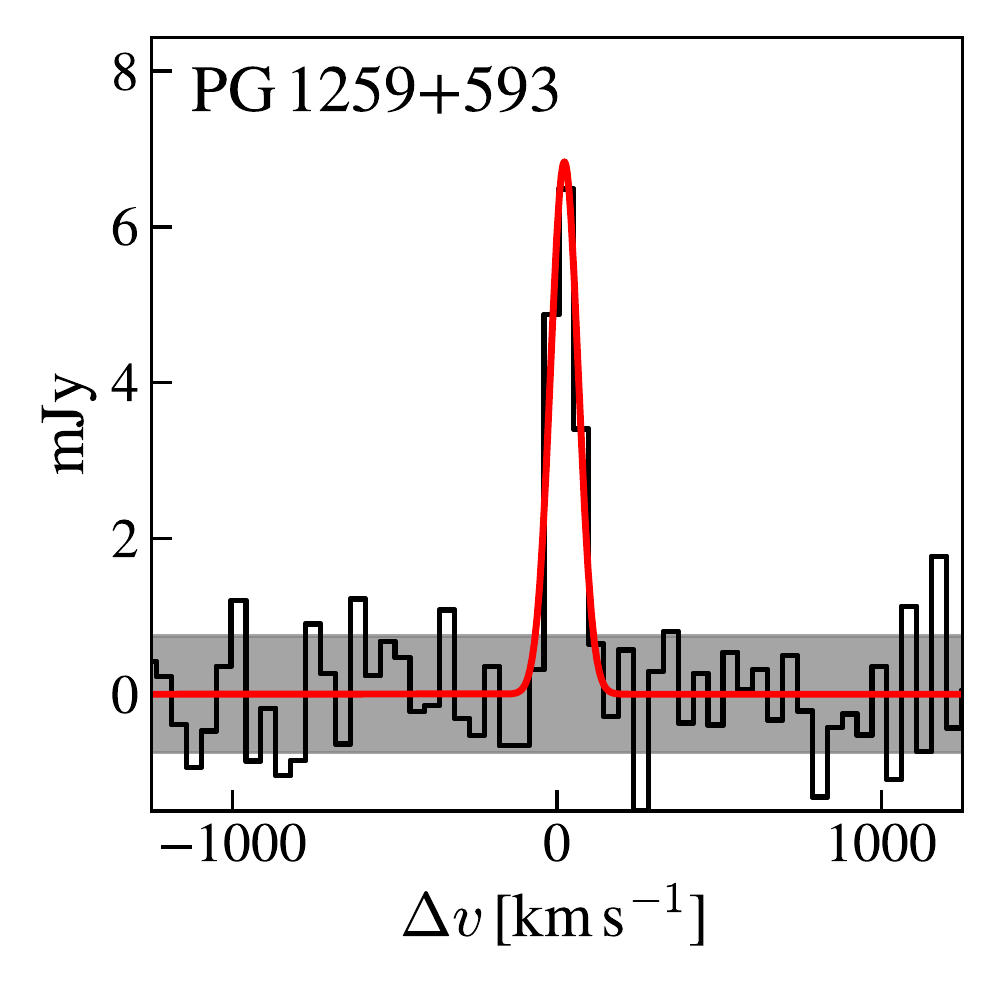} 
\includegraphics[width=0.49\columnwidth]{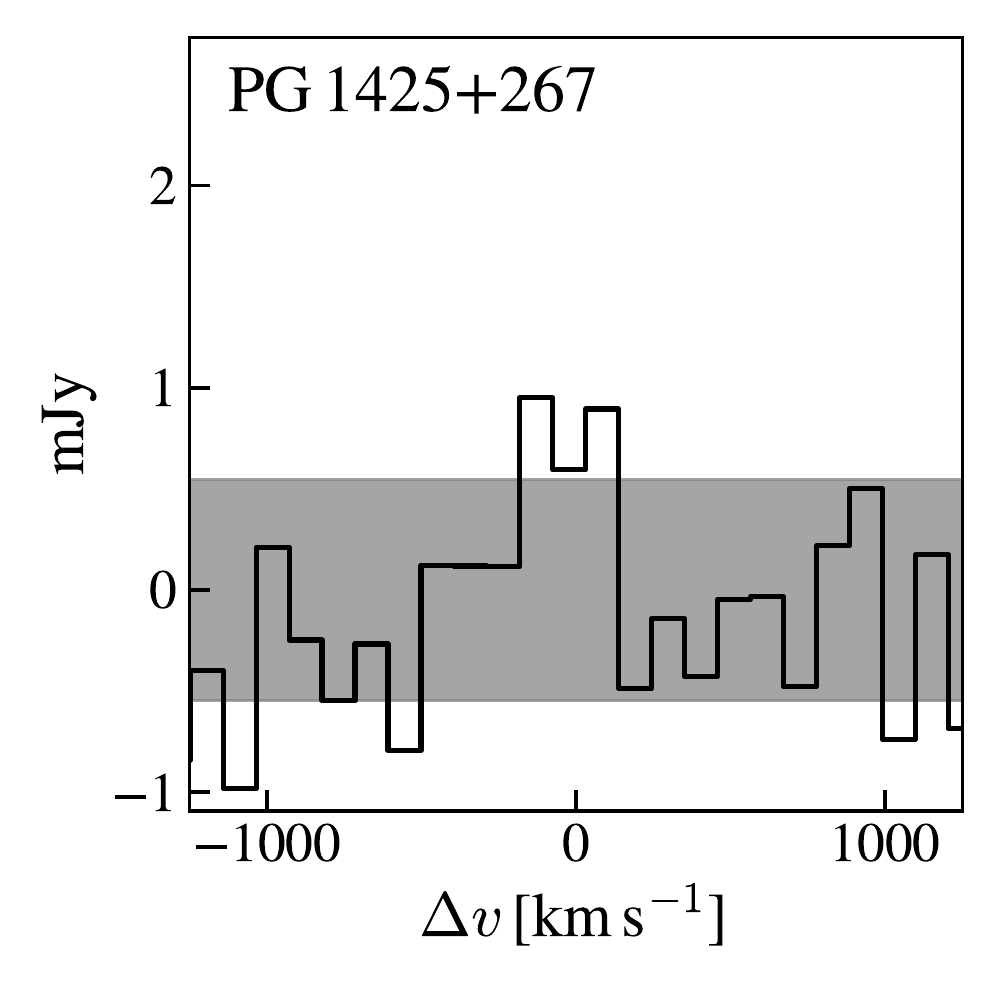}\\
\includegraphics[width=0.49\columnwidth]{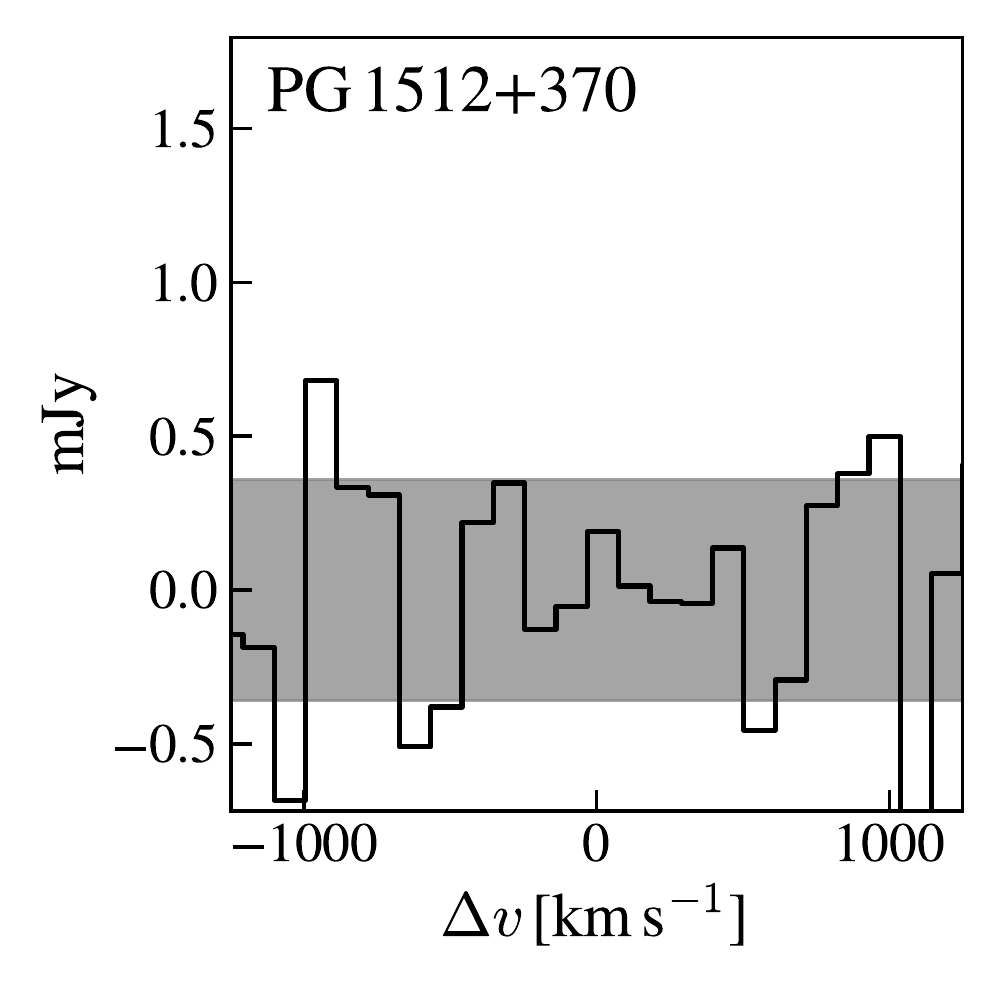}
\includegraphics[width=0.49\columnwidth]{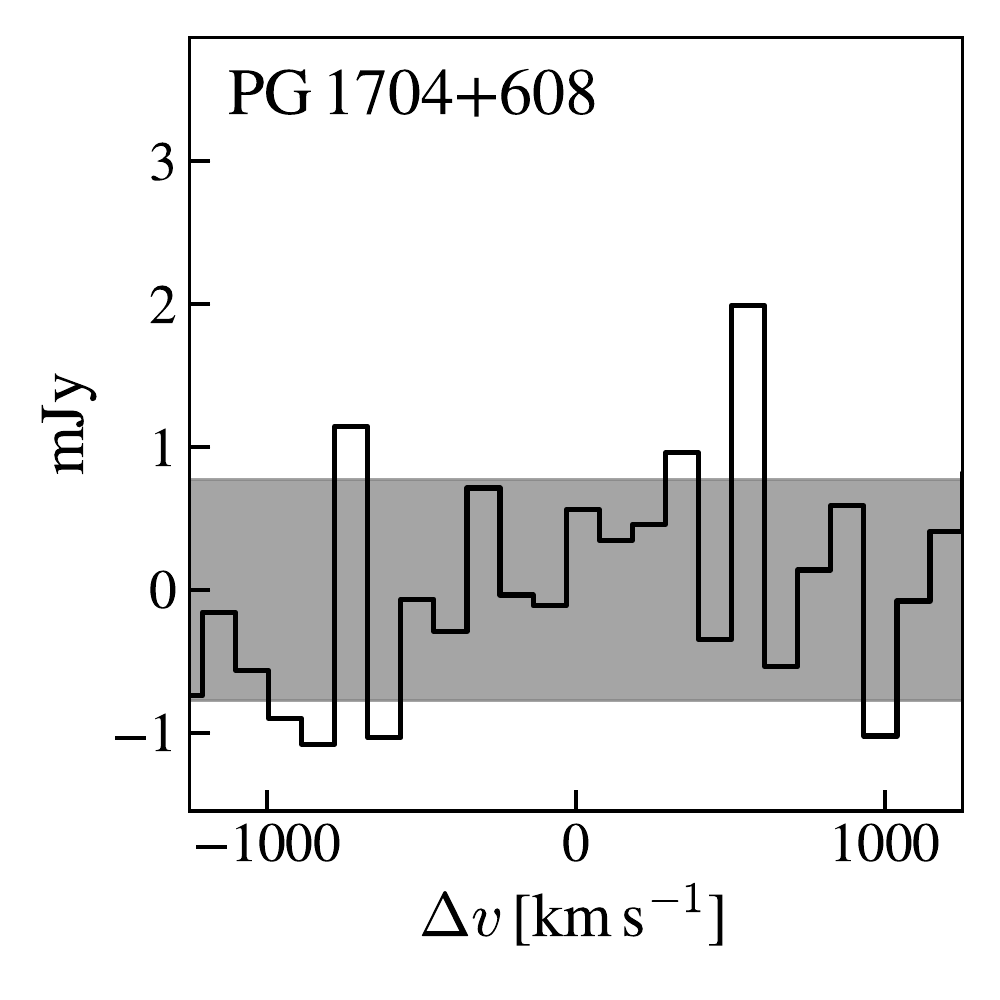}
\includegraphics[width=0.49\columnwidth]{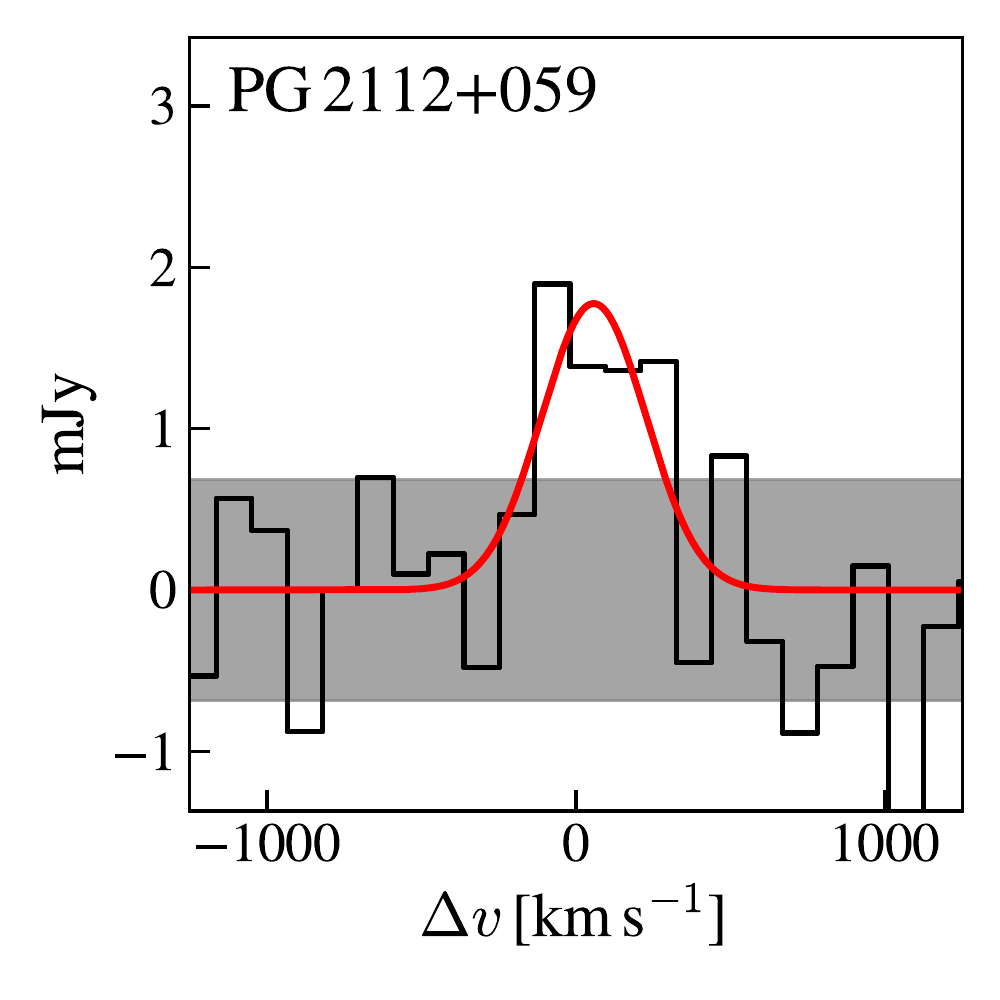}
\includegraphics[width=0.49\columnwidth]{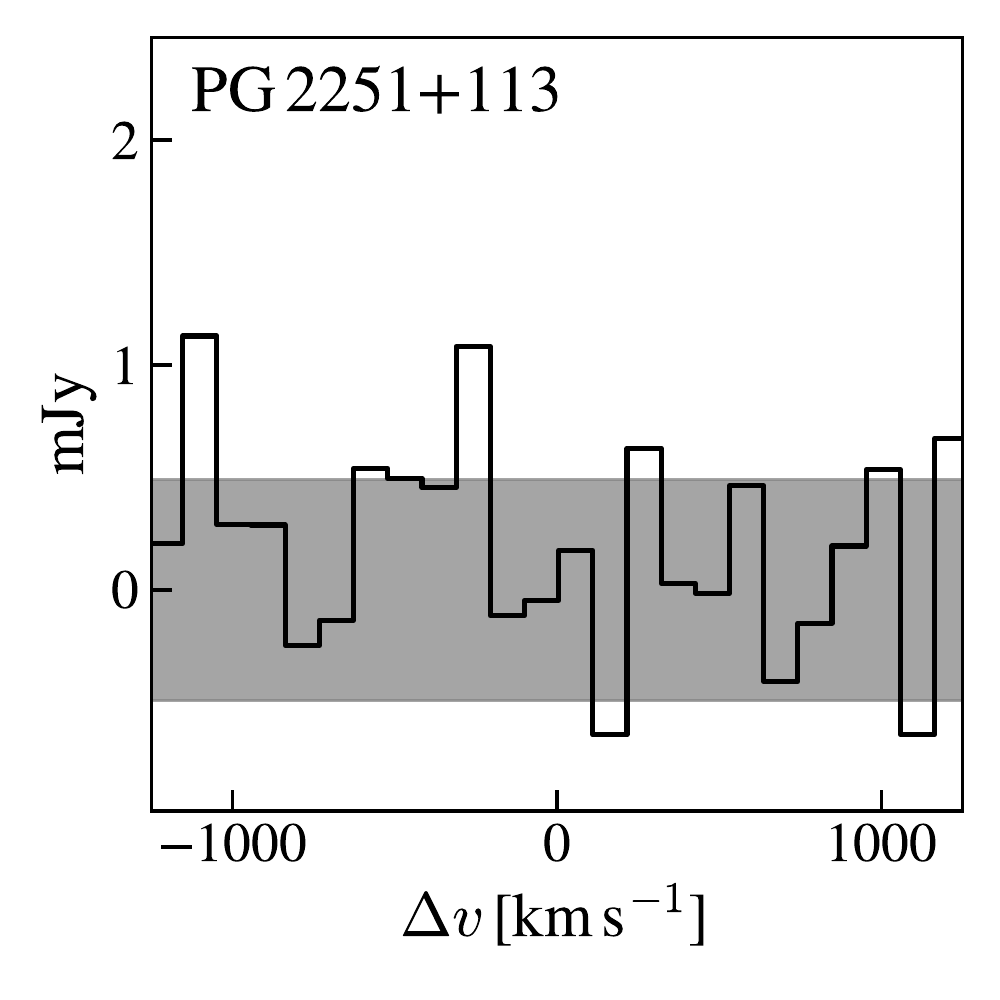}
\caption{\label{fig:CO_spectra} Spectra at restframe $\sim 1.3\,$mm of the eight PG quasar host galaxies. The red curve corresponds to the best-fit Gaussian model. The grey shade highlights the spectrum $\pm1\sigma$ level, with $\sigma$ representing the rms of emission line-free channels. In the cases where no CO(2--1) emission is detected, the spectrum is centered at the expected line location following the redshift derived from optical light.}
\end{figure*}

\begin{figure*}
\centering
\includegraphics[width=0.49\columnwidth]{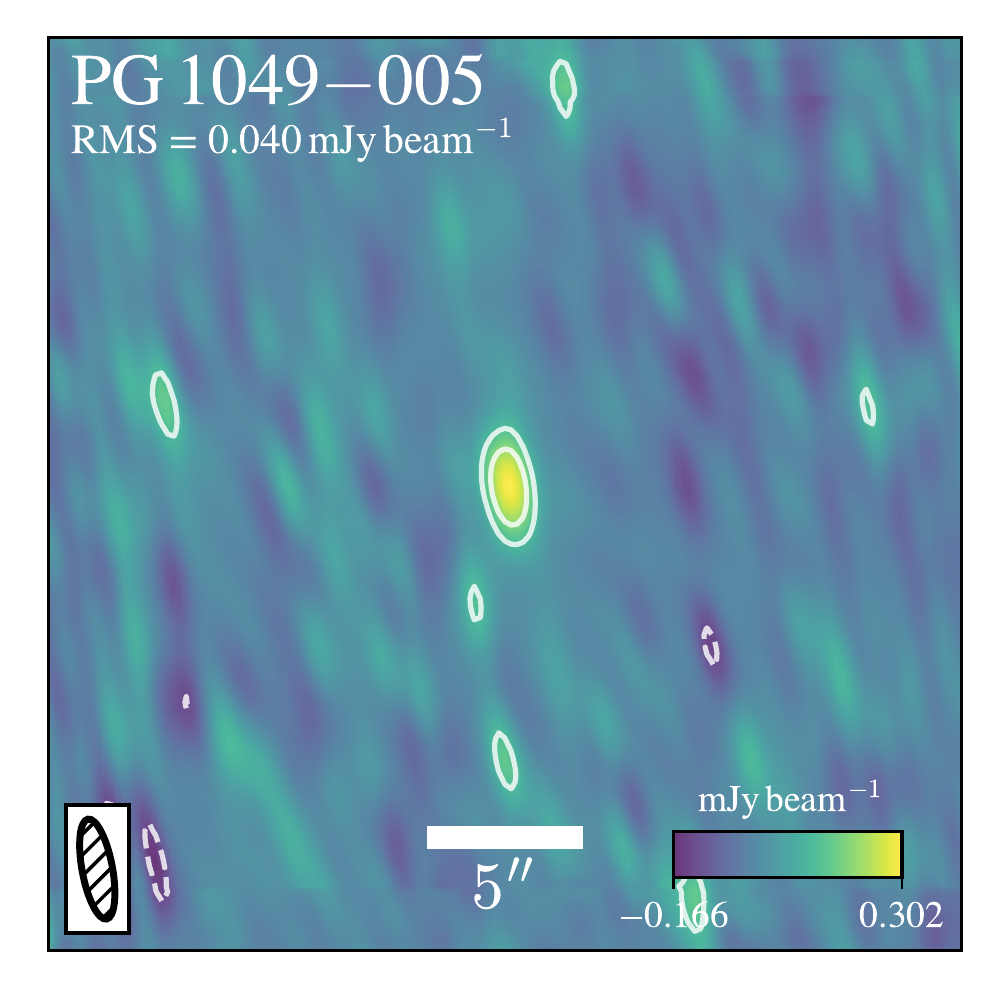}
\includegraphics[width=0.49\columnwidth]{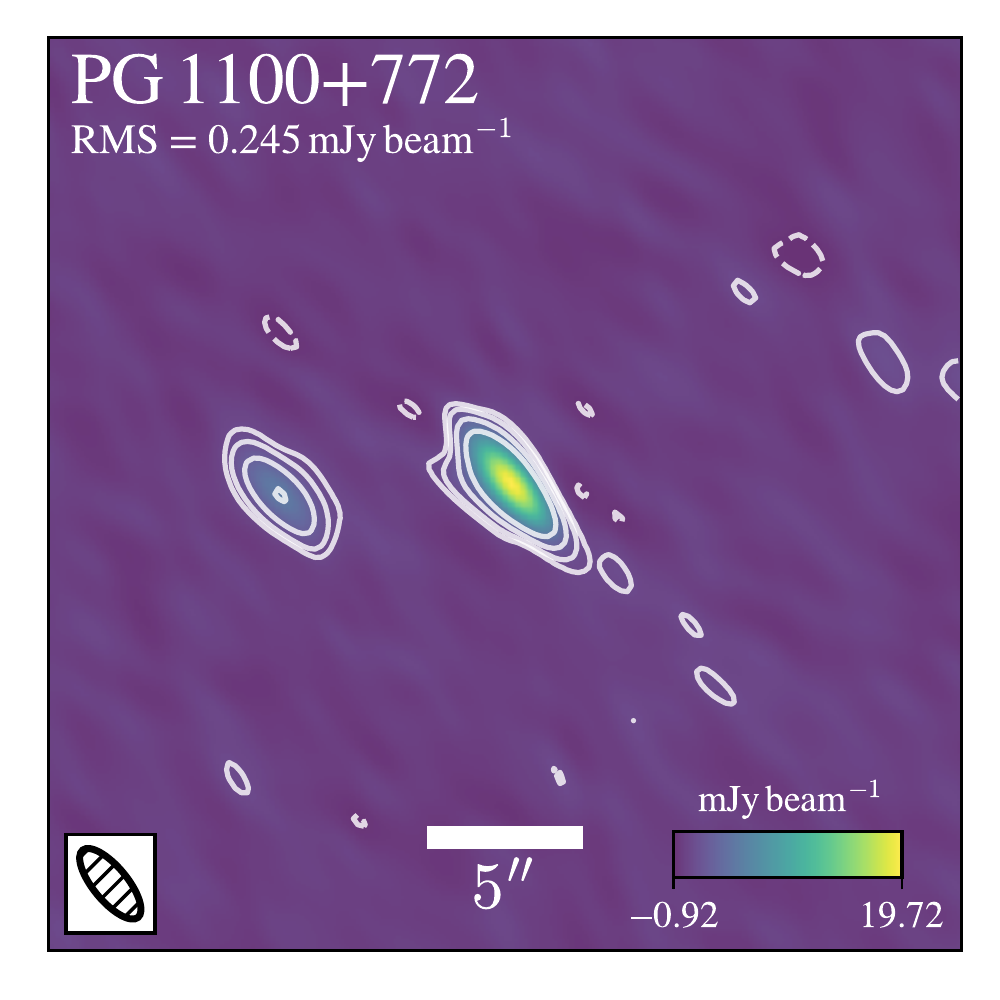}  
\includegraphics[width=0.49\columnwidth]{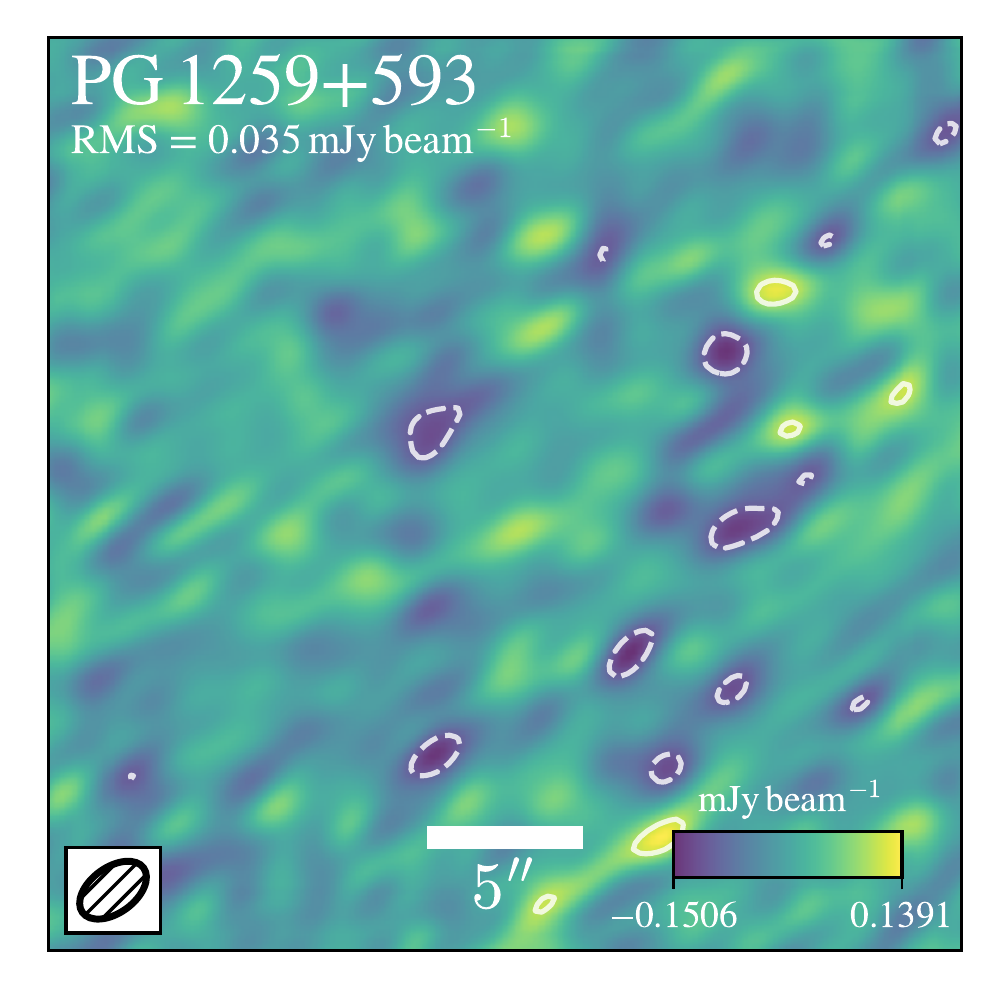} 
\includegraphics[width=0.49\columnwidth]{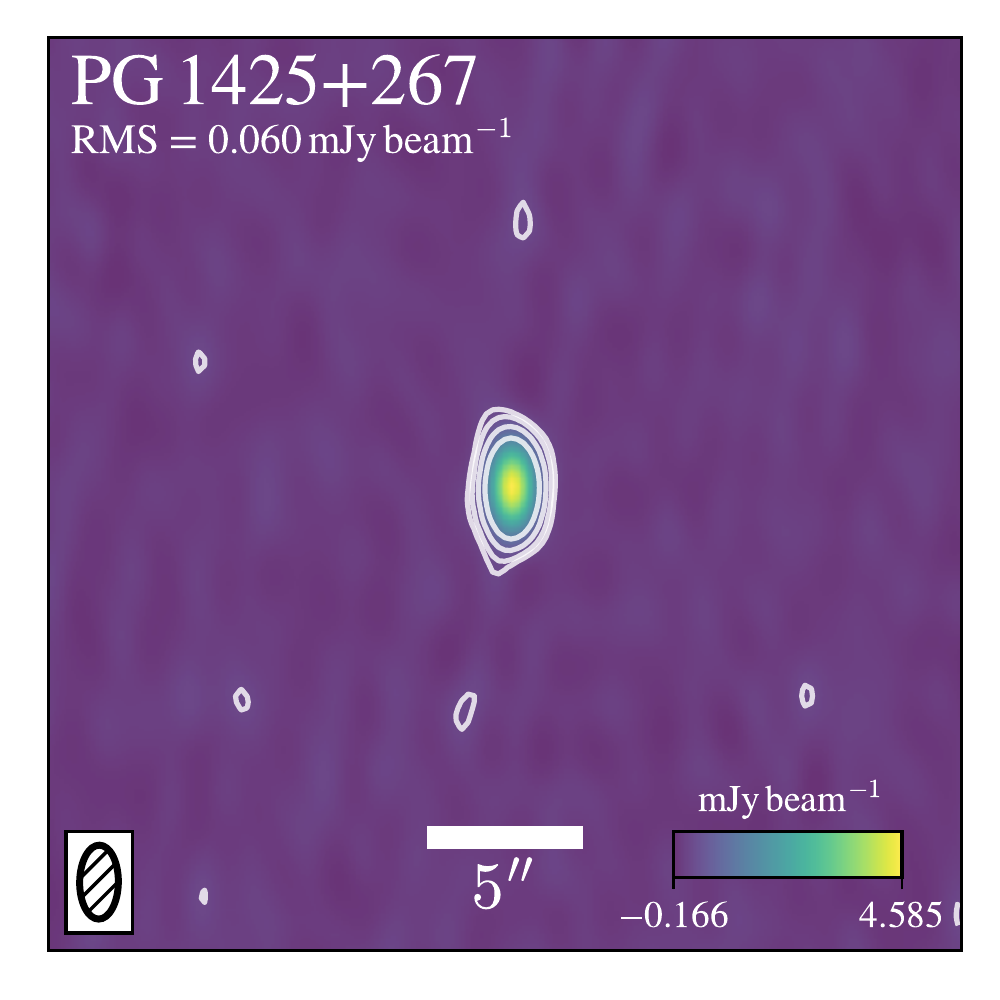}\\
\includegraphics[width=0.49\columnwidth]{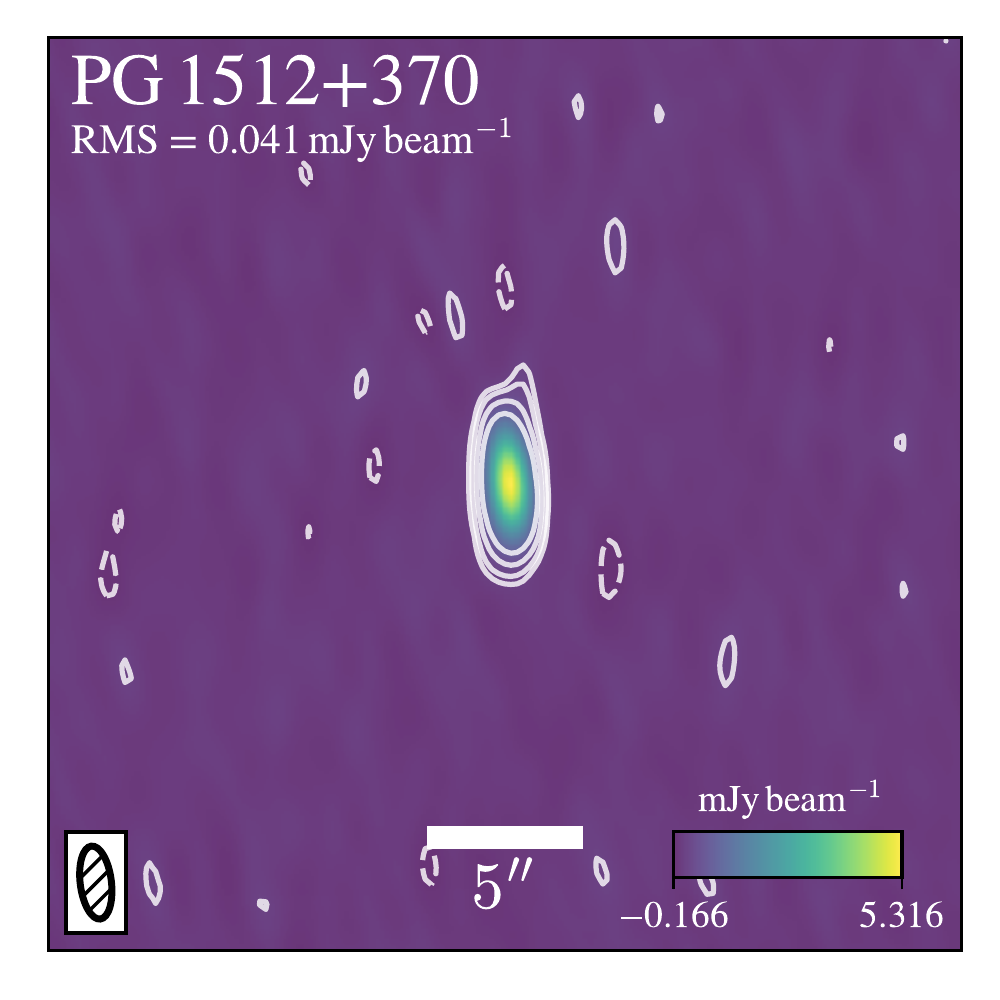}
\includegraphics[width=0.49\columnwidth]{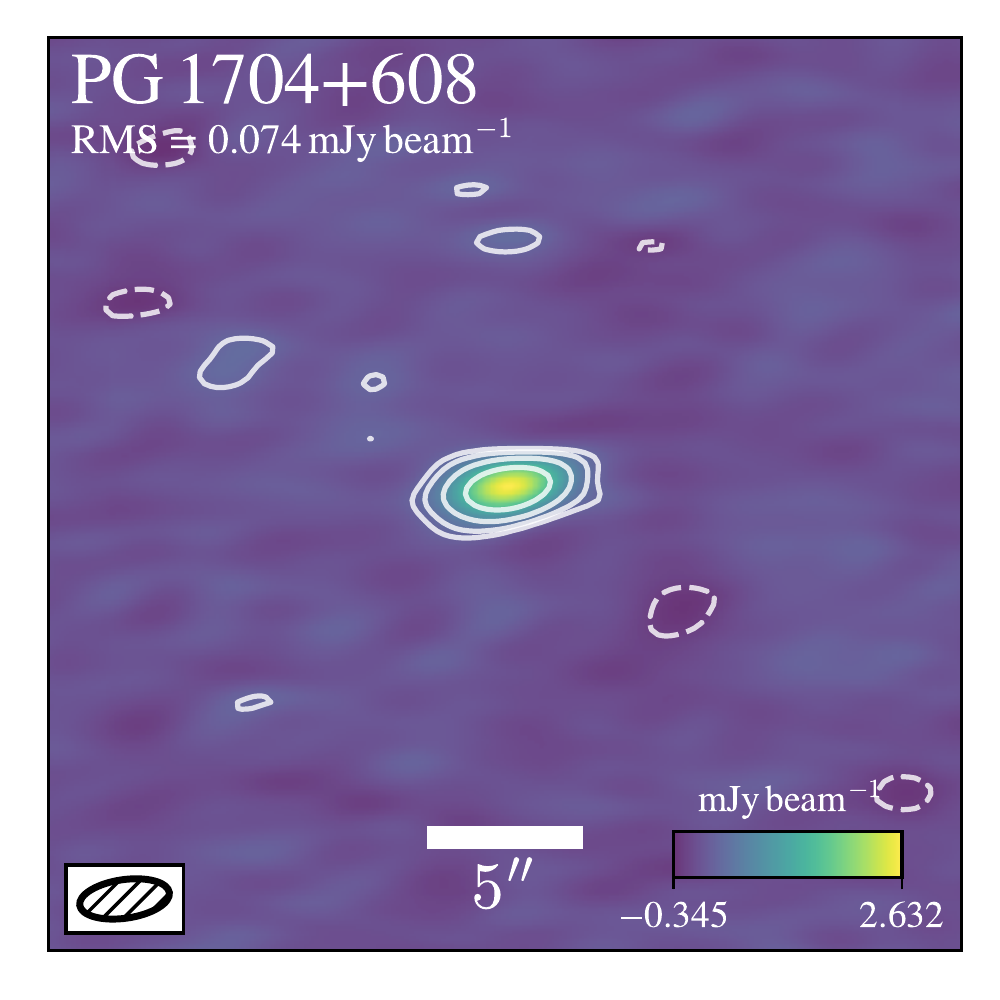}
\includegraphics[width=0.49\columnwidth]{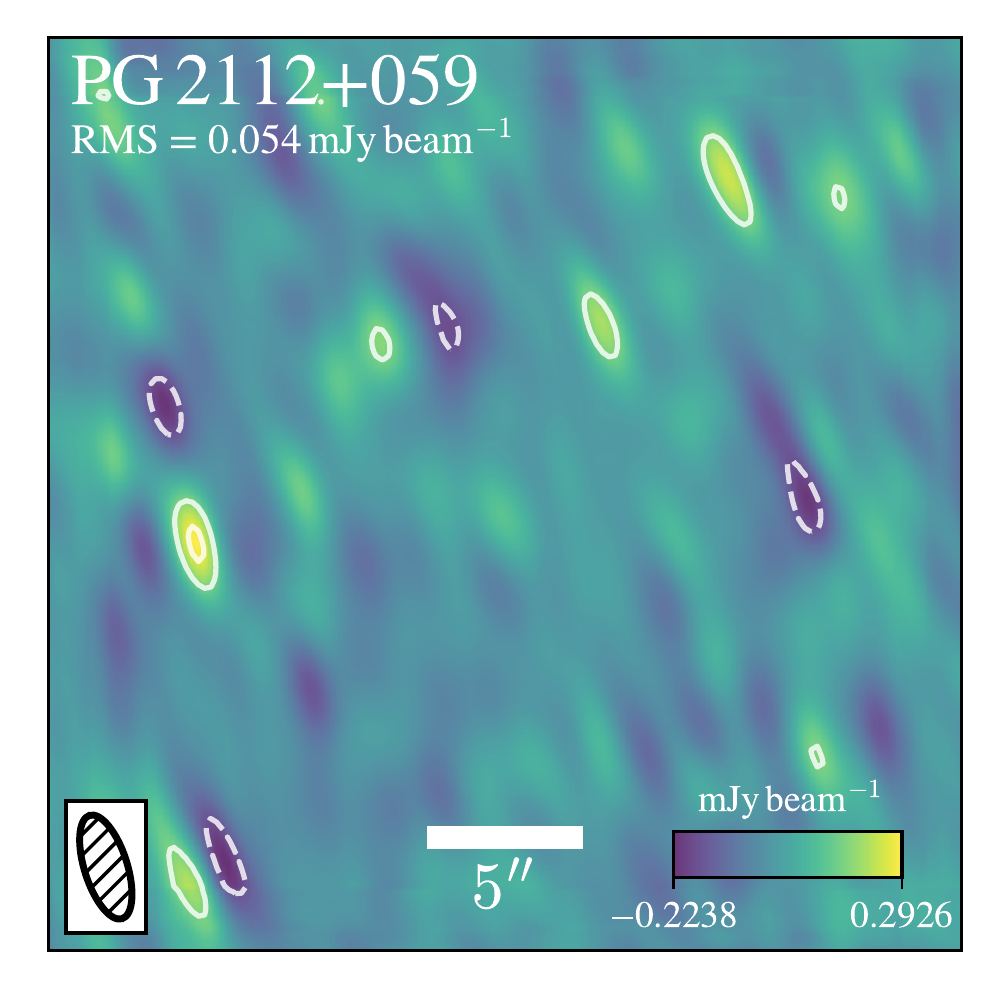}
\includegraphics[width=0.49\columnwidth]{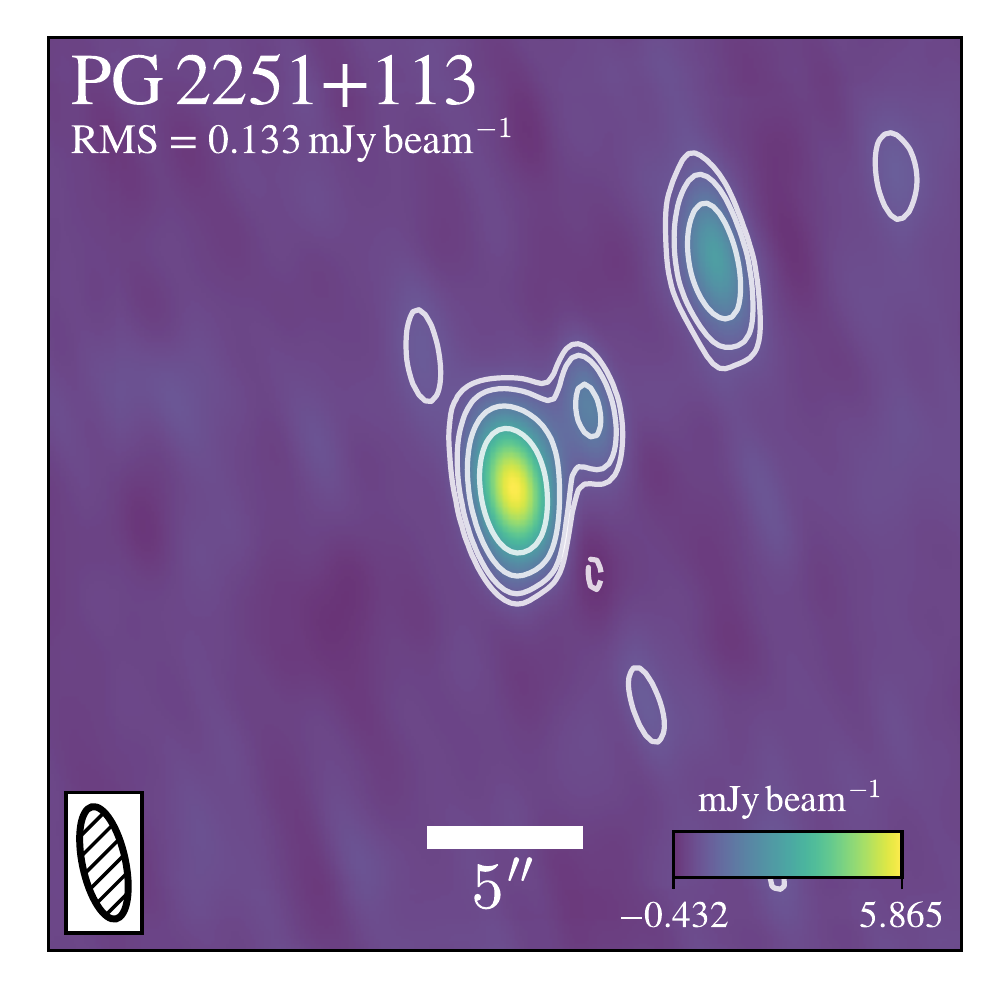}
\caption{\label{fig:cont_img} Continuum images at restframe $\sim 1.3\,$mm for the PG quasar hosts. The contours indicate $-3$ (dashed), 3, 5, 10, 20$\sigma$ levels, with $\sigma$ being the RMS value of the source-free pixels and indicated in each map.  The synthesized beam is shown in the lower left corner of each image.}
\end{figure*}

To estimate the CO(2--1) velocity integrated intensity, first, we get a spectrum by spatially collapsing the data cube within a beam-sized region centered at the expected target location (Figure~\ref{fig:CO_spectra}).  This aperture selection was set after exploring different values aiming to maximize the CO line S/N detection,  and roughly corresponds to the host galaxy stellar component size reported by \citet{Zhao2021}. Then, we fit the line emission by simply using a Gaussian function deriving the line velocity integrated intensity ($S_{\rm CO} \Delta v$), center ($\nu_{\rm obs}$) and FWHM.  We quantify the total continuum flux density ($S_{\rm cont}$) by simply fitting a two-dimensional Gaussian model to the continuum image (Figure~\ref{fig:cont_img}).  We found a good agreement between $S_{\rm cont}$ and the continuum flux density values estimated by summing the individual pixel values within an aperture of $20\arcsec$ (similar to that used for the Herschel data) centered on the continuum peak location (average ratio $\sim 1.08$,  scatter $\sim 0.23\,$dex).  In both cases, we estimate the uncertainties by re-sampling the data in 1000 iterations using the rms of the residuals as the noise level. The 1$\sigma$ errors are obtained from the distributions build for each parameter. In the case of non-detection, the CO(2--1) emission line and continuum 3$\sigma$ upper limits were calculated as

\begin{equation}  
\label{eq:uplim_CO}
S_{\rm CO} \Delta v < 3\, \sigma_{\rm CO} \, \delta v \, \sqrt{N_{\rm ch} \frac{N_{\rm A}}{N_{\rm B}}},
\end{equation}

\noindent and

\begin{equation}
\label{eq:uplim_cont}
S_{\rm cont} < 3\, \sigma_{\rm cont} \, \sqrt{\frac{N_{\rm A}}{N_{\rm B}}}
\end{equation}

\noindent respectively, where $\sigma_{\rm CO}$ is the RMS level of the data cube, $\delta v$ is the velocity resolution, $N_{\rm ch}$ is the number of channels sampling the CO line FWHM (assumed to be 400\,km\,s$^{-1}$ given the stellar masses of our targets and the $M_\star$--FWHM trend reported by \citealt{Shangguan2020} for the local PG quasars), $N_{\rm A}$ is the aperture size in pixel units, and $N_{\rm B}$ is the number of pixels for the synthesized beam.  We assume $N_{\rm A} = 3 N_{\rm B}$ for estimating the CO flux density upper limits \citep{Hainline2004,Michiyama2021}, while $N_{\rm A} = 20\arcsec$ for the continuum.   We list the estimates (or upper limits) for each source in Table~\ref{tab:CO_prop}. 

The CO(2--1) line luminosity is calculated following \citep{SV2005}

\begin{equation}
\label{eq:LCO}
L^\prime_{\rm CO(2-1)} = 3.25 \times 10^7\,\frac{S_{\rm CO(2-1)} \Delta v\, D^2_{L}}{\nu_{\rm obs}^{2}\,(1+z)^{3}}\,\,\, {\rm [K\,km\,s^{-1}\,pc^2]},
\end{equation}

\noindent where $S_{\rm CO(2-1)} \Delta v$ is in units of Jy\,km\,s$^{-1}$, $D_{L}$ is the luminosity distance in Mpc, $\nu_{\rm obs}$ is the observed frequency of the line in GHz, and $z$ is the redshift. We estimate the luminosity of the CO(1--0) line transition ($L^\prime_{\rm CO(1-0)}$) by adopting the median luminosity ratio $r_{21} \equiv L^\prime_{\rm CO(2-1)} / L^\prime_{\rm CO(1-0)} = 0.62$ found by \citet{Shangguan2020} for PG quasars at $z < 0.3$.\footnote{For the BASS AGNs \citet{Lamperti2020} report an average $r_{21} = 0.72$ value, with a median absolute deviation of 0.17.  For the CARS sample,  we use the CO(1--0) and CO(2--1) velocity integrated intensity values provided by \citet{Bertram2007} for 25 sources and estimate a median $r_{21} = 0. 61^{+0.34}_{-0.17}$ value. } The molecular gas masses are estimated under the assumption of $\alpha_{\rm CO} = 3.1$\,$M_\odot$\,(K\,km\,s$^{-1}$\,pc$^2$)$^{-1}$, as suggested by the comparison of $L^\prime_{\rm CO(1-0)}$ and the dust-based gas masses for the $z < 0.3$ PG quasars \citep{Shangguan2020},  and in consistency with the recommendation by \citet{Sandstrom2013} for nearby star-forming galaxies.

\begin{figure*}
\centering
\includegraphics[width=0.95\columnwidth]{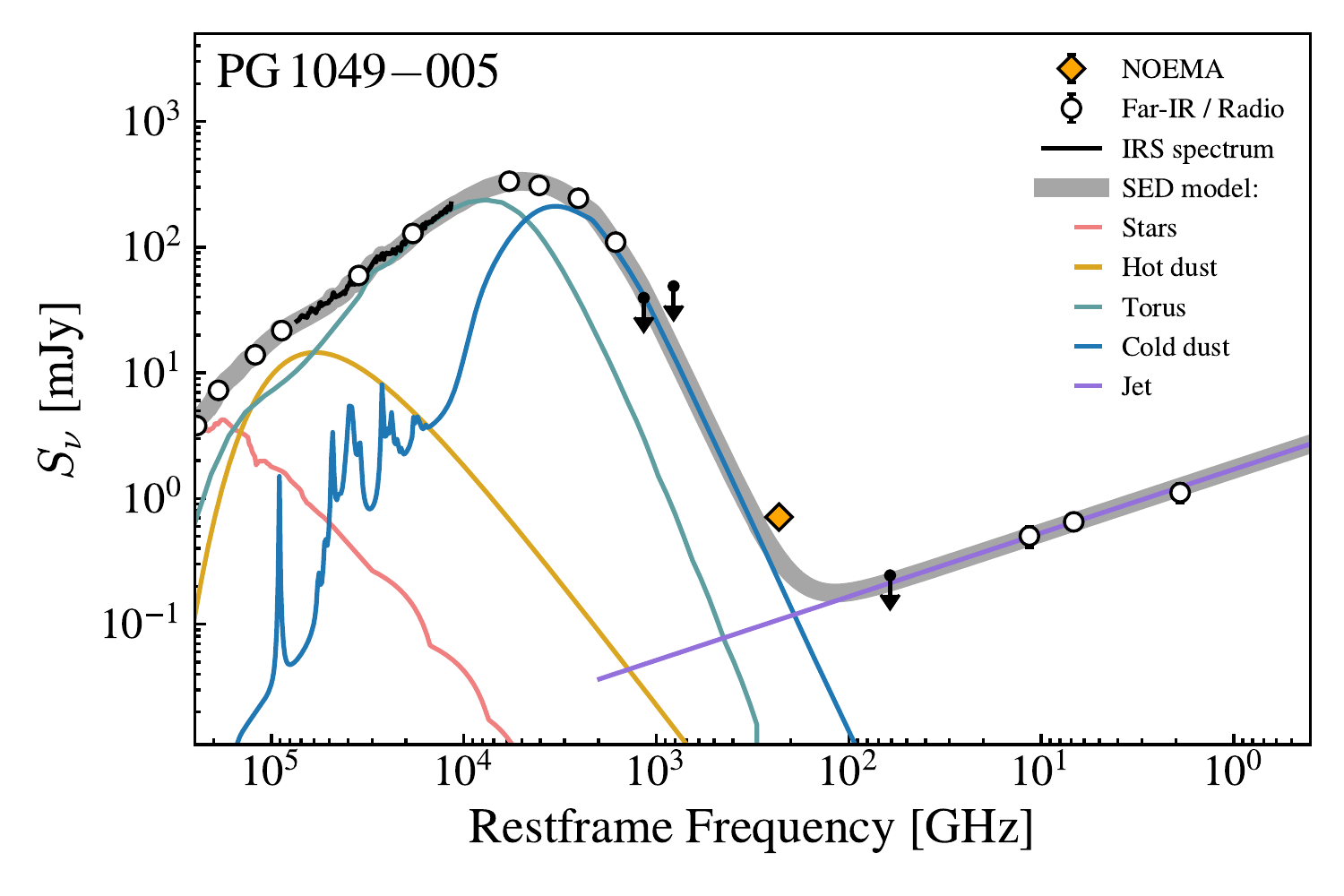}
\includegraphics[width=0.95\columnwidth]{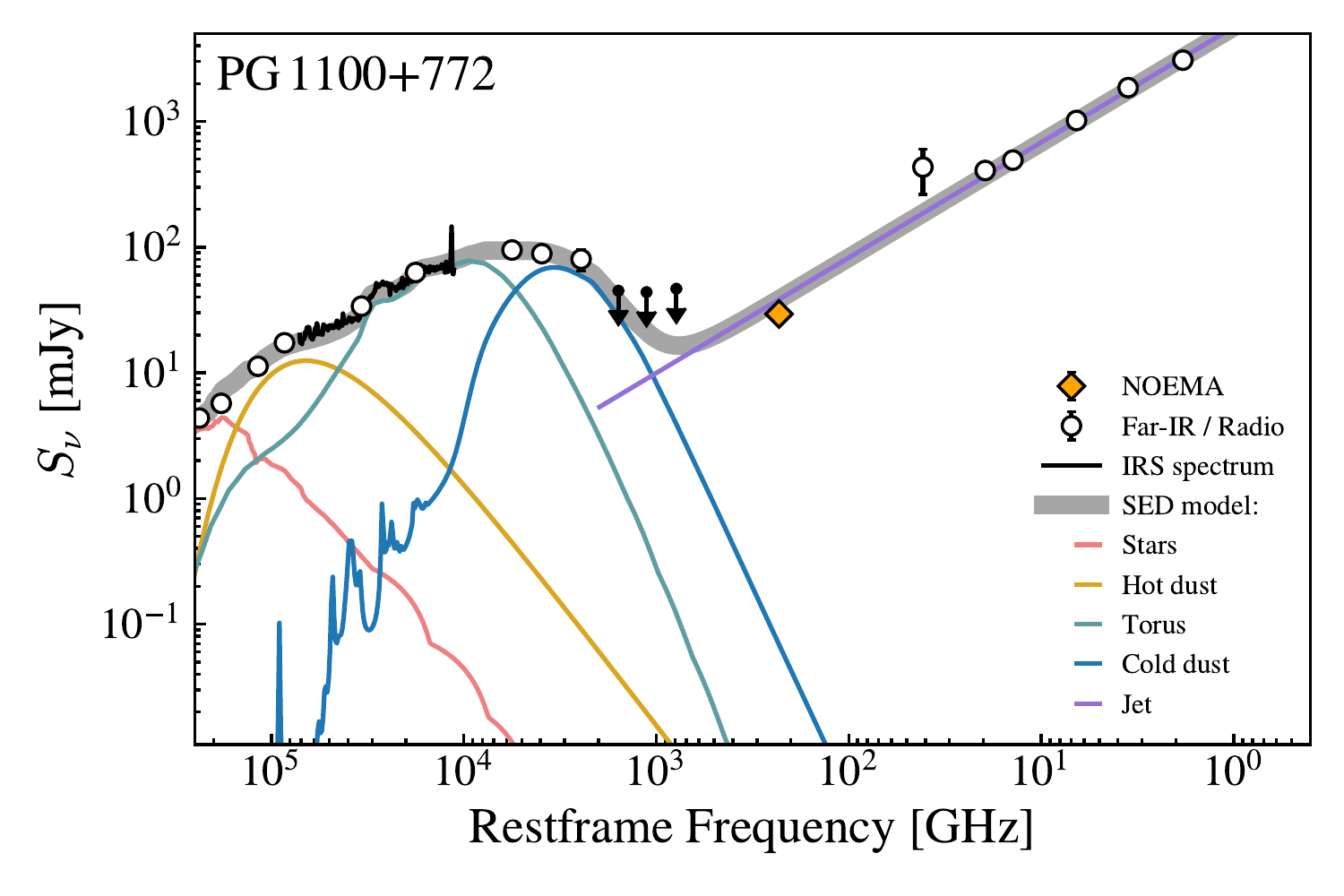}\\  
\includegraphics[width=0.95\columnwidth]{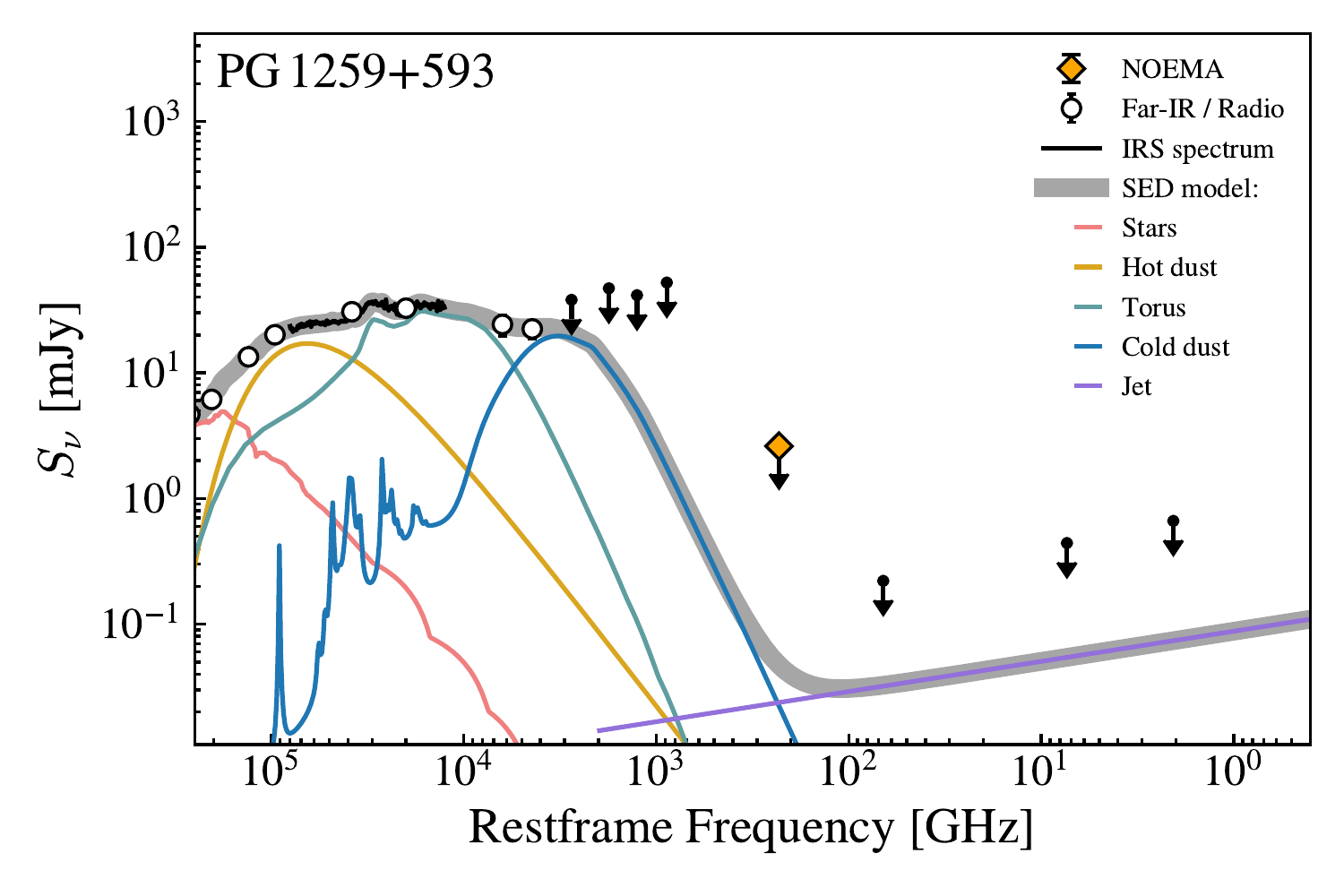} 
\includegraphics[width=0.95\columnwidth]{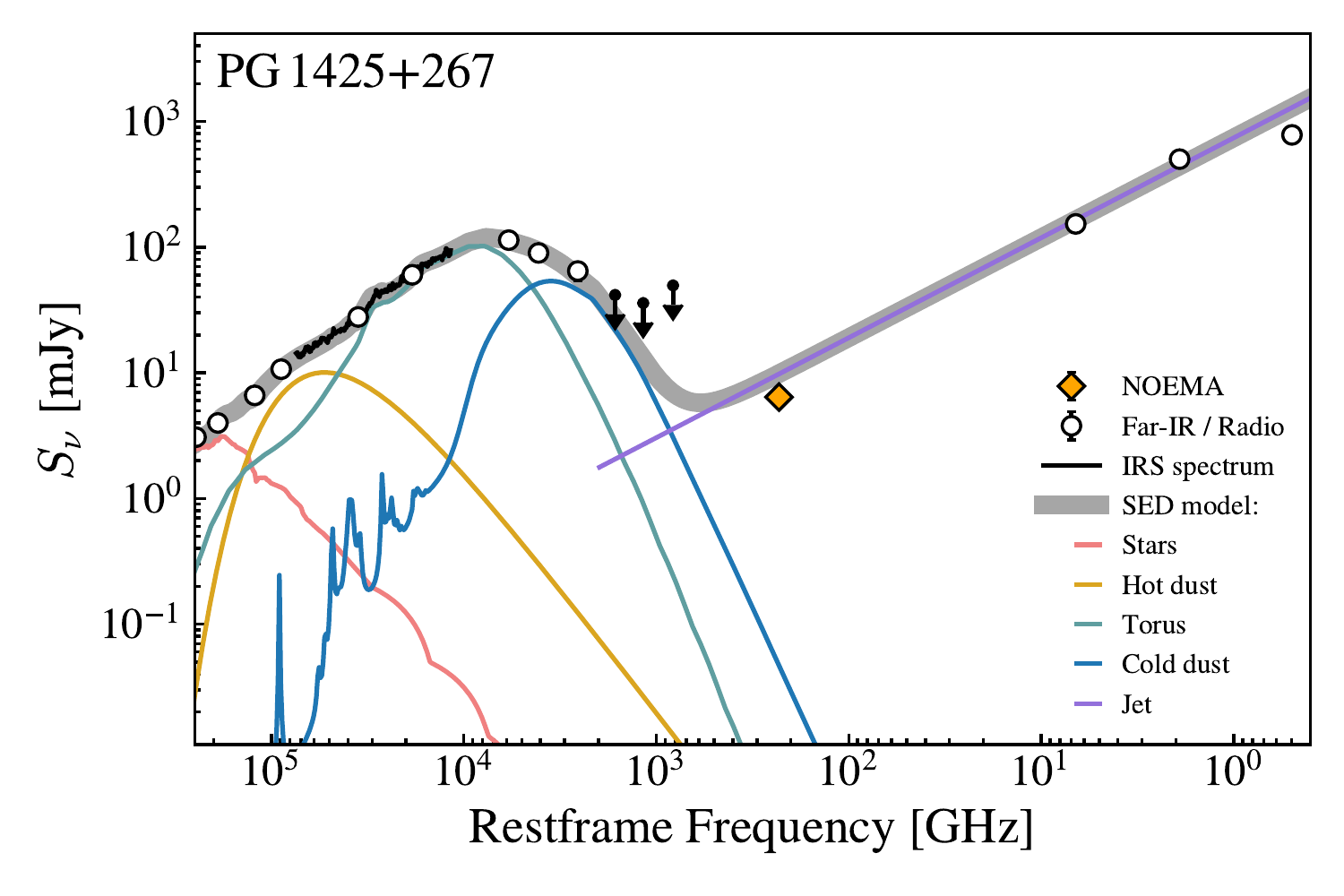}\\
\includegraphics[width=0.95\columnwidth]{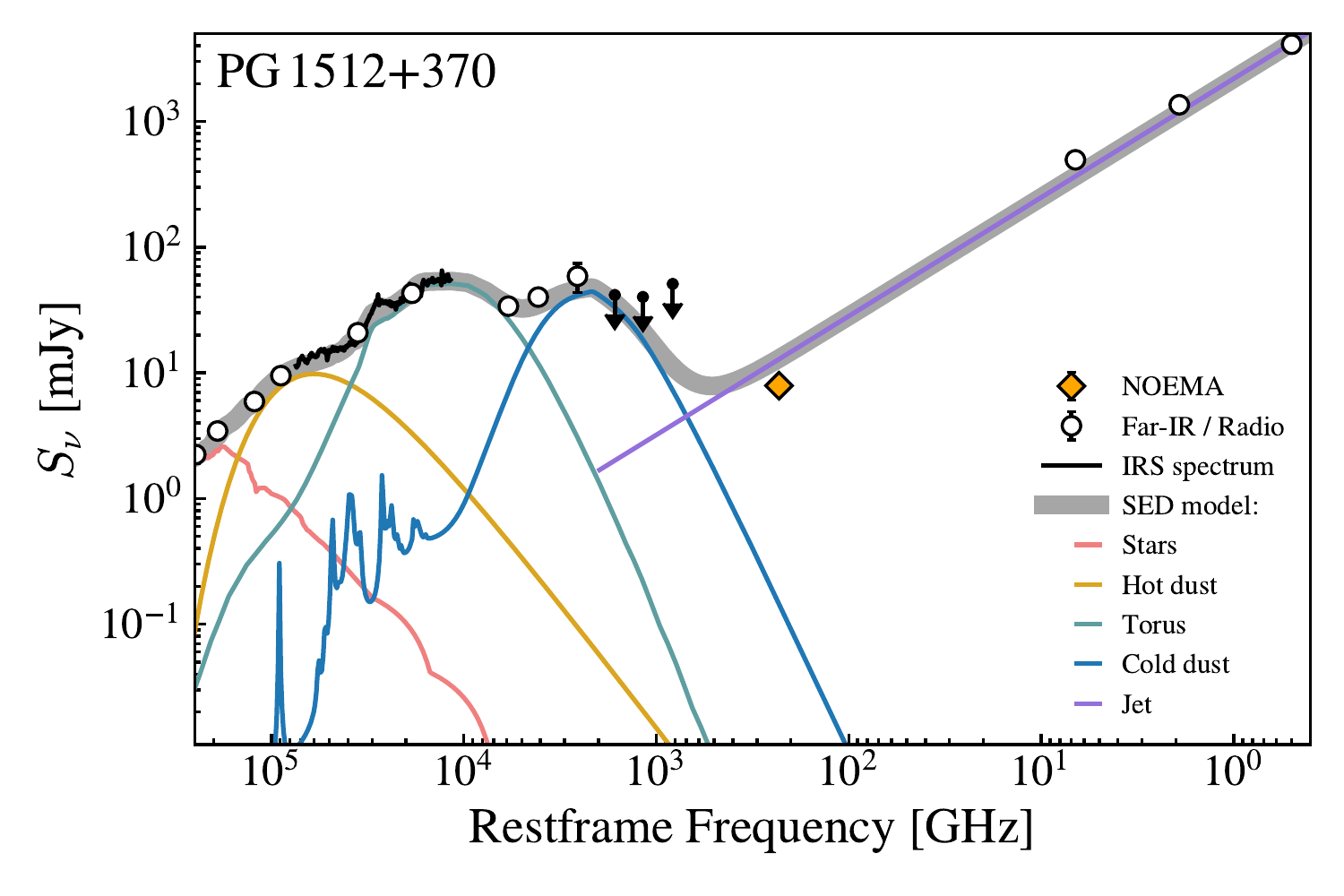}
\includegraphics[width=0.95\columnwidth]{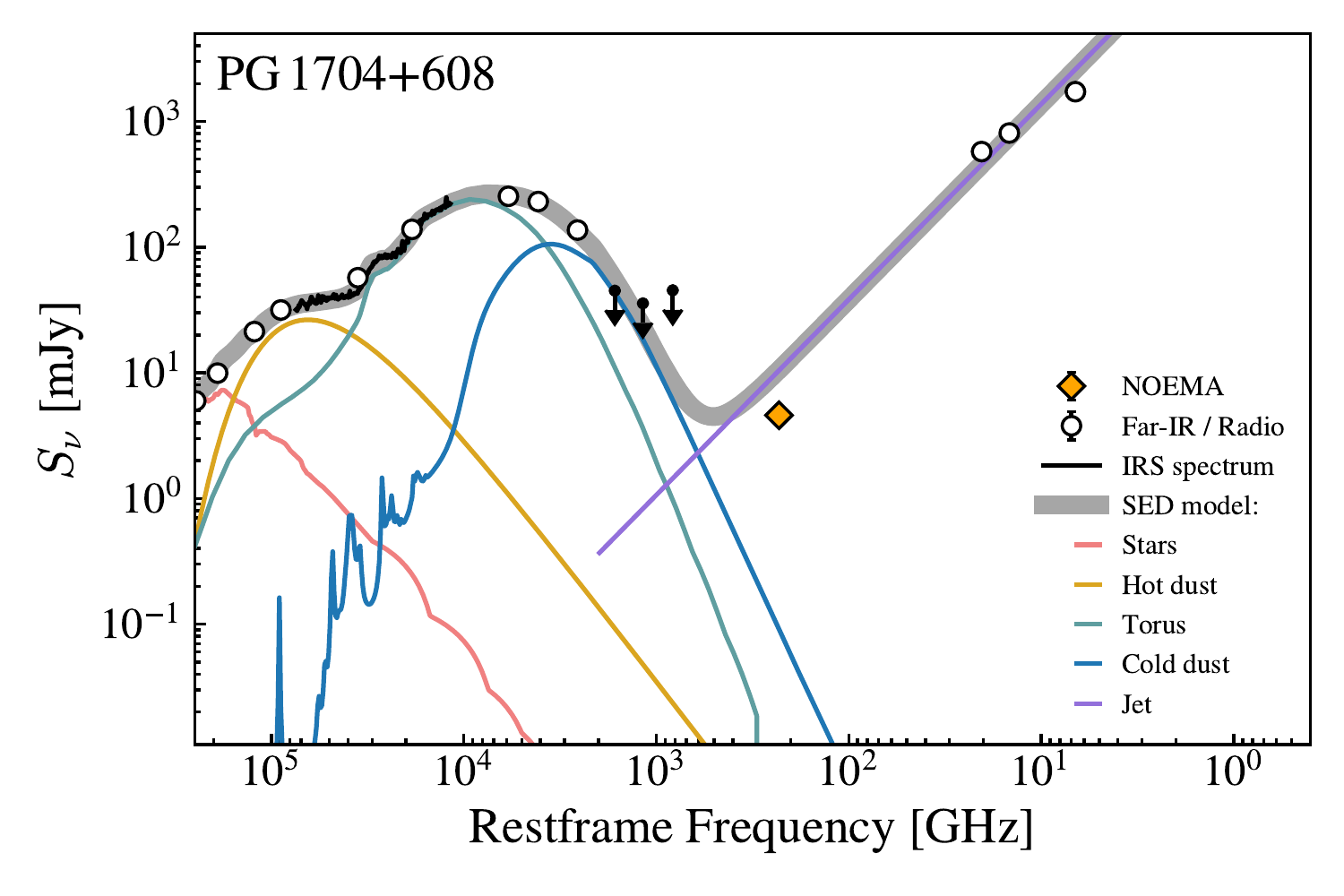}\\
\includegraphics[width=0.95\columnwidth]{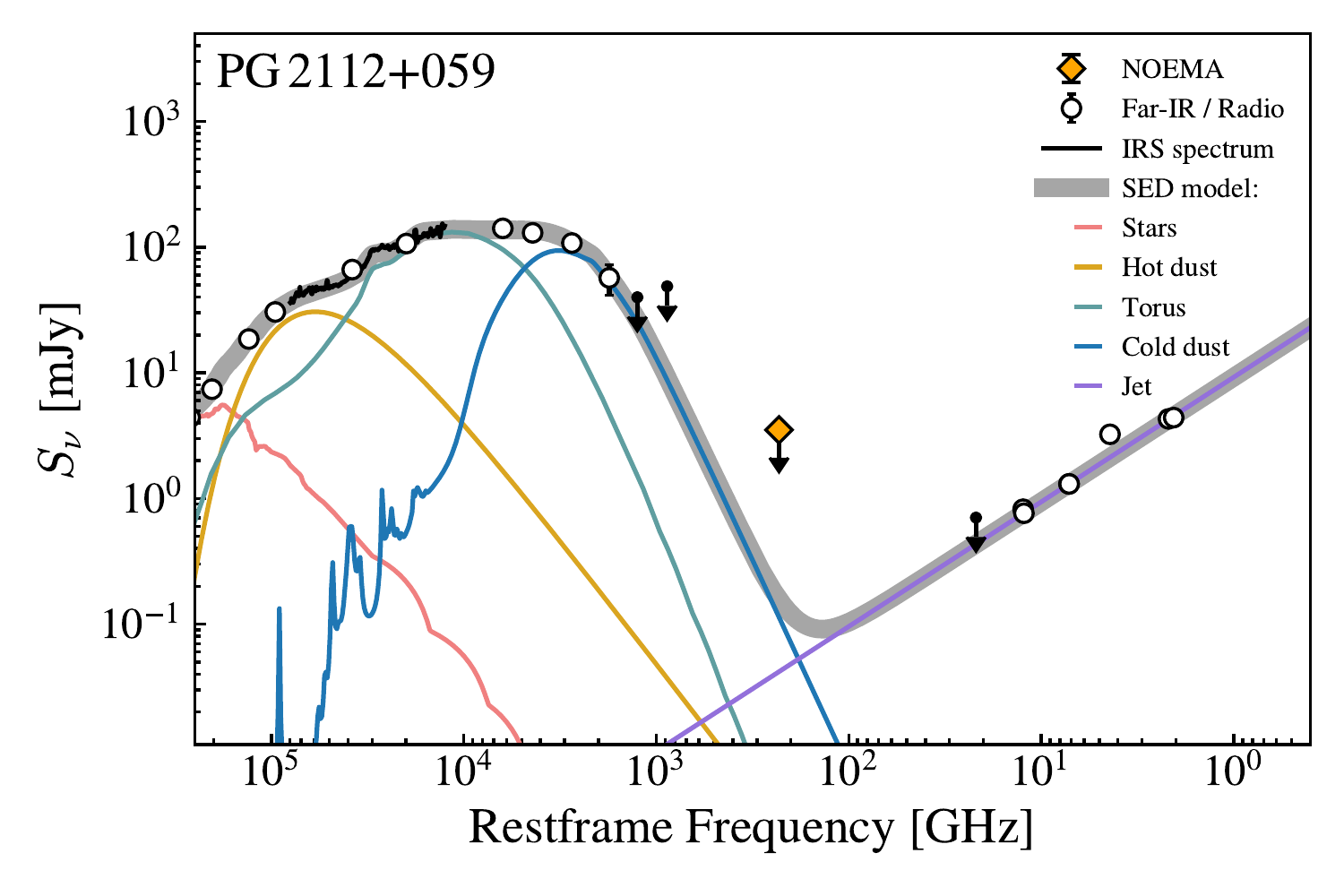}
\includegraphics[width=0.95\columnwidth]{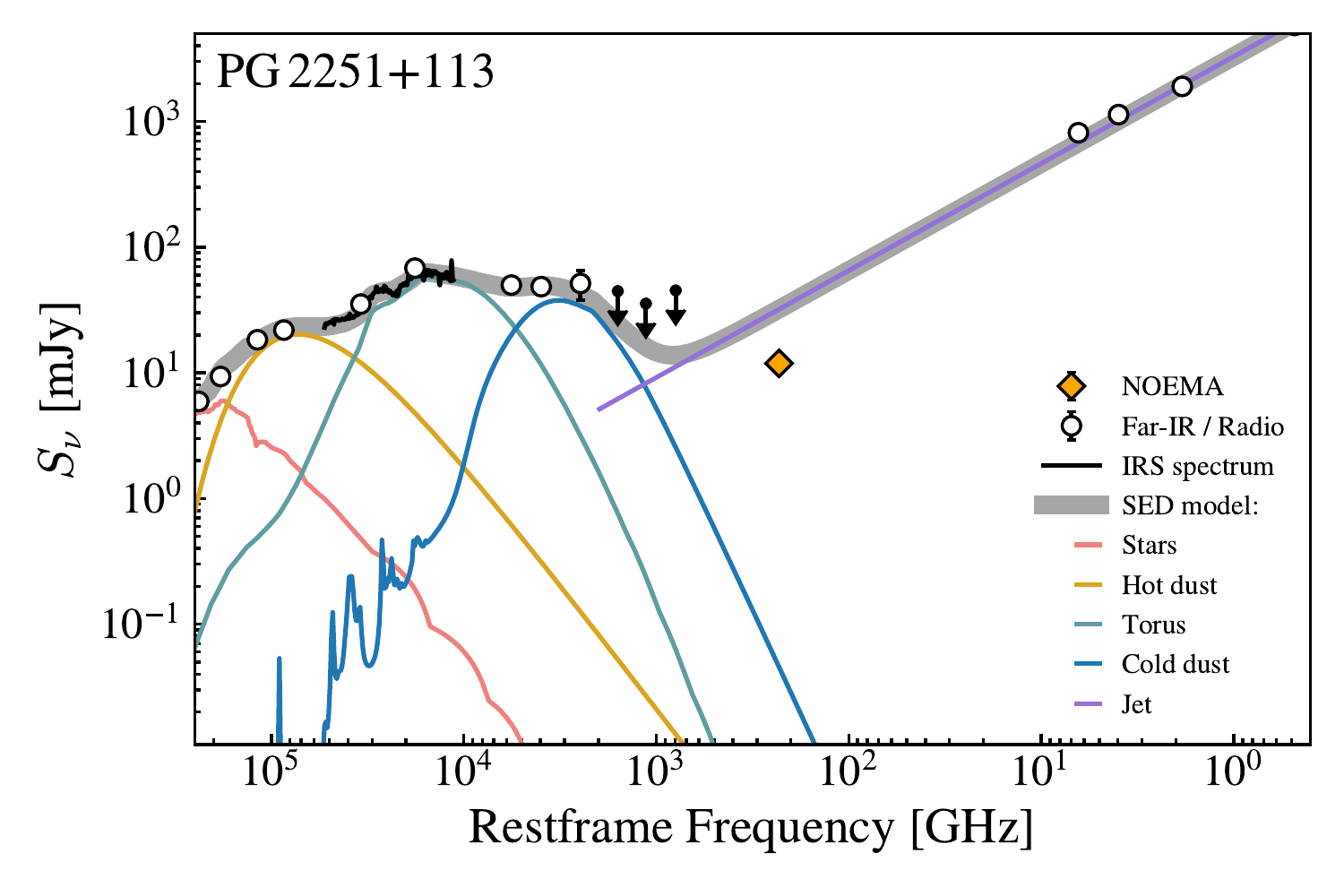}
\caption{\label{fig:far-IR_SED} Far-IR SED of the eight PG quasar host galaxies. The orange diamond indicates the NOEMA $\sim2\,$mm continuum. The open circles show the IR data from 2MASS, WISE, and Herschel collated in \citet[see Section~\ref{sec:farIR_SED}]{Shangguan2018}, along with the GHz radio data taken from the literature. Upper limits are indicated by downward arrows (Appendix~\ref{sec:AppA}).  The black solid line represents the available IRS spectrum in the $\sim 5-40\mu$m wavelength range. The far-IR SED model is in grey band, and it is composed by five sub-components: starlight (red), hot dust (yellow), AGN torus (green), cold dust (blue) and jet (purple; Section~\ref{sec:SED_modeling}).}
\end{figure*}

\subsection{Far-IR SED modeling and IR-based SFRs}
\label{sec:SED_modeling}

We use our $S_{\rm cont}$ estimates to complement the far-IR SED (Section~\ref{sec:farIR_SED}) and update the AGN-decontaminated IR luminosity values previously estimated by \citet{Shangguan2018} for our targets. Our $S_{\rm cont}$ measurements provide key information for constraining the targets far-IR SED flux density at low frequencies.

\citet{Shangguan2018} modeled the far-IR SED by adopting physically motivated emission components for the AGN dusty torus, host galaxy cold dust, and starlight.  The torus emission is modeled by using the CLUMPY model \citep{Nenkova2008a, Nenkova2008b}, plus a complementary blackbody component that accounts for emission from the torus very hot dust \citep{Zhuang2018}. The far-IR models adopt a fixed dust emissivity index ($\beta$) within the \citet{DraineLi2007} models (roughly $\beta \approx 2.08$). The far-IR modeling also handle a synchrotron emission component, but its addition to the SED is optional to the radio data availability. The synchrotron emission is modeled as a power-law spectrum with fixed spectral index ($S_\nu \propto \nu^{\alpha_{\rm R}}$).  A 5\,Gyr stellar population with a \citet{Kroupa2001} IMF from \citet{Bruzual2003} is assumed for the starlight component. We refer to \citet{Shangguan2018}, for more details.  The SED data and models are presented in Figure~\ref{fig:far-IR_SED}.
 
We compute the IR luminosity by integrating the host galaxy far-IR SED sub-component corresponding to the cold dust emission and over $8-1000\,\mu$m (Table~\ref{tab:CO_prop}). To estimate SFRs, we use \citet{Kennicutt1998b}'s calibration, re-normalized to a \citet{Kroupa2001} IMF,
\begin{equation}
\label{eq:SFR}
{\rm SFR}(M_\odot\,{\rm yr}^{-1}) = 3 \times 10^{-44} \, L_{\rm IR}({\rm erg\,s}^{-1}).
\end{equation}
\noindent We note that the systematic uncertainties of the far-IR SED model decomposition dominate the SFR errors. \citet{Xie2021} found that the IR-based SFRs correlate with the estimates from the mid-IR neon emission lines within $\sim 0.2-0.3\,$dex scatter for the $z < 0.3$ PG quasar hosts.\footnote{While \citet{Xie2021} analyzed the full sample of $z < 0.5$ PG quasars with Herschel data \citep{Shangguan2018}, the most distant host galaxy detected in mid-IR neon emission is at $z \approx 0.32$ (PG\,2251+113).} Hereafter, we assume this scatter value as the 1$\sigma$ error of our SFR estimates.

\section{Results} 
\label{sec:res}

Our main goal is to study whether the AGN may affect the cold molecular gas content in a sample of the most powerful unobscured quasars observed at low-$z$. From the sub-sample of 8 quasars presented in this work, we detect the CO(2--1) emission coming from 3 host galaxies (Figure~\ref{fig:CO_spectra}).  Those 3 systems tend to have the less massive BHs and higher Eddington ratios among the sample of 8 objects. Considering both the NOEMA and ACA campaigns, we detect line emission in 36 out of 48 host galaxies. All 17 disk-like host galaxies are detected in CO, while we detect CO emission in 7 out of 9 merger systems, and in 5 ellipticals from a total of 12 (5 non-detections from this campaign). We detect the CO emission in 7/10 host galaxies with uncertain morphology classification.  Using the qualitative radio spectra classification for the PG quasars \citep{Shangguan2018}, we find that 33/38 radio-quiet systems are detected in CO, 3 out of 4 PG quasars with flat radio spectra present line detection, and no CO emission was detected in systems with steep radio spectra (6 in total).

Thanks to the large frequency coverage of NOEMA, we detect the restframe $\sim1.3\,$mm continuum in 6 systems (Figure~\ref{fig:cont_img}). The modeling of the global far-IR SEDs (Figure~\ref{fig:far-IR_SED}) suggests that the restframe $\sim1.3\,$mm continuum can be associated with synchrotron emission from a jet in 5 sources. We now proceed to analyze the data in detail.

\begin{figure}
\centering
\includegraphics[width=1.0\columnwidth]{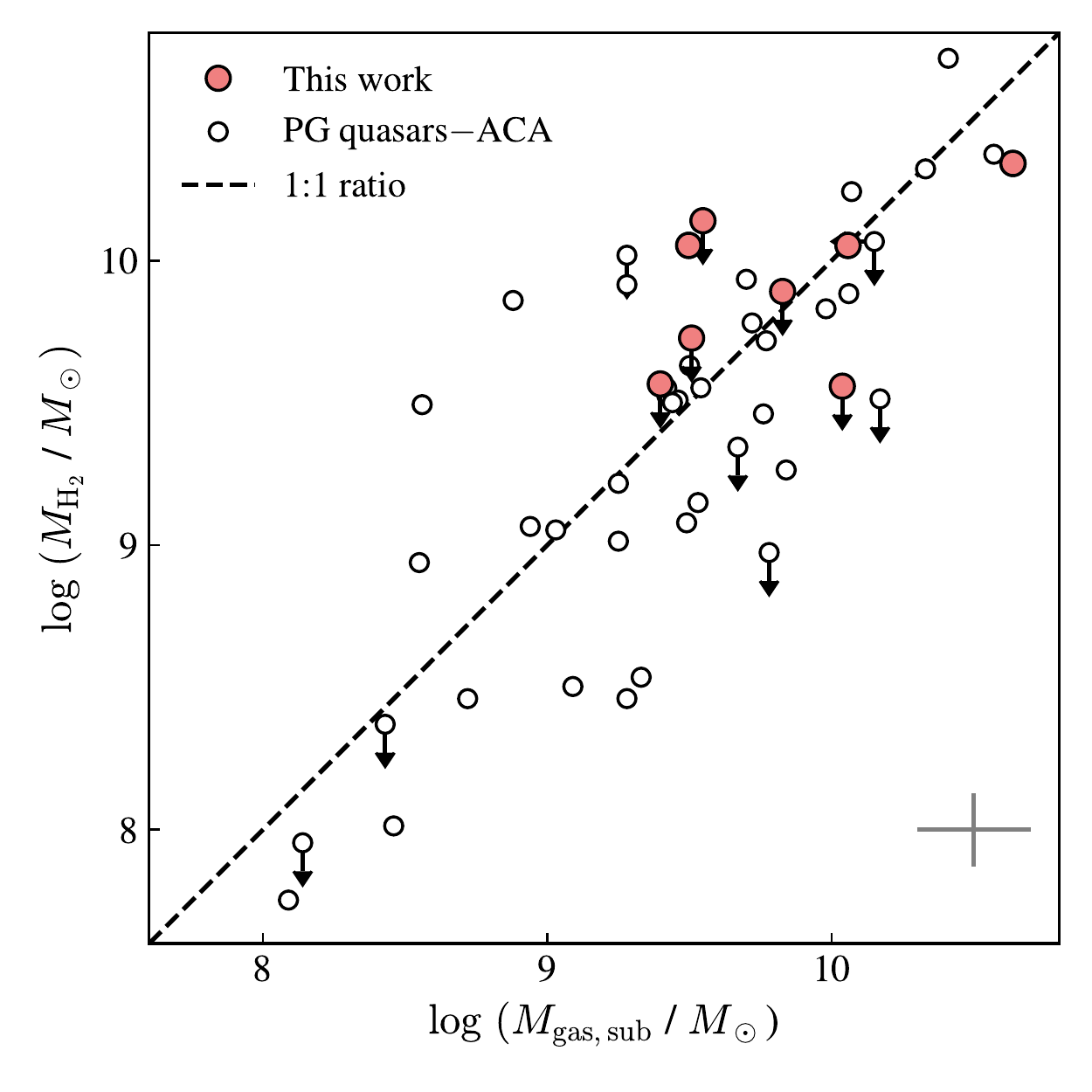}
\caption{\label{fig:Herschel_comp} CO-based molecular gas mass compared to the total gas mass predicted from cold dust emission and subtracted for extended atomic gas (H\,{\sc i}) content.  In the bottom right corner we present the typical $1\sigma$ uncertainty of the data. The arrows denote to $3\sigma$ upper limits. }
\end{figure}

\subsection{Total Gas Mass Content Estimated from Cold Dust Emission}
\label{sec:Dust_comp}

The far-IR SED of galaxies can be used to predict the dust masses,  which are closely related to their total gas content \citep{Draine2007,Leroy2011}.
\citet{Shangguan2018} provided total gas mass ($M_{\rm gas}$) estimates for the PG quasar host galaxies, finding that those closely correlate with direct measurements from CO emission plus atomic gas (H\,{\sc i}) for the closer systems. Here, we extend their analysis by comparing both the molecular gas masses estimated from the CO(2--1) emission-line luminosity and $M_{\rm gas}$ values for the $z < 0.5$ PG quasar host galaxy sample. The latter quantity is computed by converting the dust masses to $M_{\rm gas}$ using the dust-to-gas ratio values derived by \citet{Shangguan2018}, but avoiding the correction factor ($\sim 0.23$\,dex) for extended atomic hydrogen gas content ($M_{\rm gas, sub}$).  Hence, the $M_{\rm gas, sub}$ estimates correspond to the atomic plus molecular gas mass within roughly the same host galaxy area where the dust emission comes from \citep{Shangguan2018}.  The $M_{\rm gas}$ values have a systematic uncertainty of $\sim0.2\,$dex, while the CO-based estimates have a typical uncertainty of $\sim0.32\,$dex inherent to $\alpha_{\rm CO}$ ($\sim0.3\,$dex),  and $R_{21}$ ($\sim0.1\,$dex). We remind that the $L'_{\rm CO (2-1)}$ uncertainty is dominated by the flux calibration uncertainties. 

\begin{figure*}
\centering
\includegraphics[width=2.0\columnwidth]{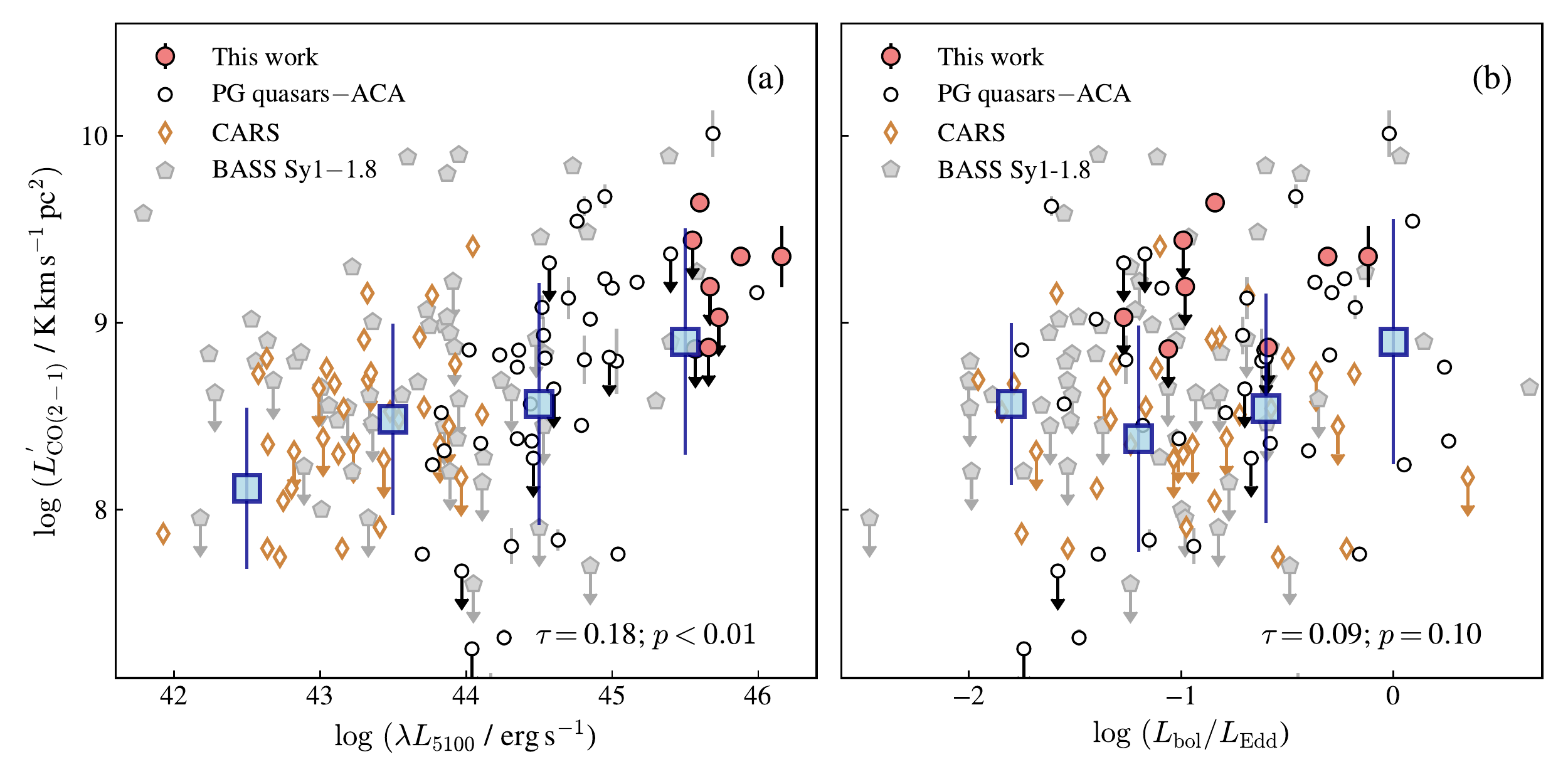}
\caption{\label{fig:LCO_L5100} The dependence of CO luminosity on (a) $\lambda L_{5100}$ and (b) Eddington ratio. In both panels, the blue squares correspond to median $L'_{\rm CO(2-1)}$ values computed in $0.6\,$dex $\lambda L_{5100}$ and $0.3\,$dex $L_{\rm bol}/L_{\rm Edd}$ bins. The arrows correspond to $3\sigma$ upper limits.}
\end{figure*}

Figure~\ref{fig:Herschel_comp} shows $M_{\rm H_2}$ against $M_{\rm gas, sub}$ for the full sample of PG quasars (NOEMA plus ACA campaigns).  Both quantities closely match, in agreement with the previous finding of \citet[their Figure 10]{Shangguan2018}.  We compute a mean $M_{\rm H_2} / M_{\rm gas,sub}$ ratio of 0.93 with scatter $\sim 0.40\,$dex. The scatter is consistent with the mass ratio systematic uncertainty ($\sim 0.38\,$dex),  but we note that it may be larger considering the $M_{\rm H_2}$ upper limits.  Among the PG quasars, the disk-like host galaxies present an average $M_{\rm H_2} / M_{\rm gas,sub} \approx 0. 95$ and scatter slightly reduced to $0.32\,$dex. $M_{\rm gas,sub}$ tends to be larger than $M_{\rm H_2}$ in mergers.  For such systems, the data average ratio and scatter is 0.80 and $0.23\,$dex, respectively; however, we caution the low statistics (7 host galaxies). We only detect the CO emission in 5 elliptical host galaxies, so we do not provide any statistical measurement. For the PG quasar hosts with uncertain morphological classification, we find an average mass ratio of $0.77$, with scatter $\sim 0.51$\,dex.  Our sample is too small to statistically characterize the sub-sample of PG quasar host galaxies presenting a flat or steep radio spectra. The radio-quiet quasars are characterized by an average $M_{\rm H_2} / M_{\rm gas,sub} \approx 0.92$, with scatter of $0.37\,$dex.  Our observations suggest low atomic gas content inside the stellar disks of the PG quasar host galaxies, but this result sensibly depends on the correction factor applied to the dust-to-gas ratio,  namely the extended H\,{\sc i} mass value subtracted to $M_{\rm gas}$.  Only 13 PG quasars hosts have H\,{\sc i} global measurements from literature, most of these host galaxies are ongoing mergers where the emission line exhibit complex profiles and suggest dynamical disturbance, or have neighboring companions (e.g., PG\,0007+106 and PG\,1119+120) inducing signal confusion \citep{Shangguan2018}.  Detailed observations resolving the H\,{\sc i} content in PG quasars hosts are needed to draw deeper conclusions.

\subsection{AGN Properties Compared to $L'_{\rm CO(2-1)}$}
\label{sec:AGN_LCO}

We compare the CO(2--1) luminosity with respect to $\lambda L_{5100}$ and the Eddington ratio in Figure~\ref{fig:LCO_L5100} for the PG,  BASS, and CARS type~1 AGNs.  We use \texttt{pymccorrelation} \citep{Privon2020} to compute the generalized Kendall's rank correlation coefficient ($\tau$) considering the censored data \citep{Isobe1986}.  We assume $p \leq 0.05$ to determine if the quantities are correlated.  We find that the CO luminosity does not correlate with the Eddington ratio with $\tau = 0.09$ and a $p$-value of $0.10$.  This is consistent with the results from the previous ACA campaign targeting the PG quasar hosts \citep{Shangguan2020b}, but inconsistent with the findings of \citet{Koss2021}, who found a tentative increase of molecular gas mass (hence, $L'_{\rm CO(2-1)}$) with Eddington ratio for the full BASS AGN sample (their Figure 15). Nevertheless, \citet{Koss2021} analyzed type~1 and type~2 AGNs, while we only analyze type~1 AGNs. We find that $L'_{\rm CO(2-1)}$ and $\lambda L_{5100}$ are weakly correlated ($\tau = 0.18$,  $p < 0.01$).  This result agrees with that reported by \citet{Shangguan2020b},  but the correlation coefficient is $\sim 2$ times lower due to the inclusion of the less luminous BASS and CARS AGNs. 

We also provide median trends by computing median values over bins of 1.0\,dex in $\lambda L_{5100}$ and 0.6\,dex in $L_{\rm bol}/L_{\rm Edd}$ using the Kaplan-Meier estimator implemented in the \texttt{Python} package \texttt{lifelines},\footnote{https://lifelines.readthedocs.io/en/latest/index.html\#} including CO(2--1) upper limits.  The median trend between $L'_{\rm CO(2-1)}$ and $L_{\rm bol}/L_{\rm Edd}$ is nearly constant,  as expected because both quantities are not correlated. In terms of $L'_{\rm CO(2-1)}$ and $\lambda L_{5100}$,  the median trend suggests an $\sim 0.4\,$dex increase of $L'_{\rm CO(2-1)}$ for AGNs with $\lambda L_{5100} \gtrsim 10^{44}$\,erg\,s$^{-1}$.  Such an increase of $L'_{\rm CO(2-1)}$ with $\lambda L_{5100}$ is not artificially produced by host galaxy mass selection bias as most of these AGNs tend to be hosted in galaxies with $M_\star \approx 10^{10.75}-10^{11.5} \, M_\odot$.  We note that, the AGNs with $\lambda L_{5100} \lesssim 10^{44}$\,erg\,s$^{-1}$ tend to be observed in the nearby Universe ($z \lesssim 0.1$), while the more luminous quasars with $\lambda L_{5100} \approx 10^{46}$\,erg\,s$^{-1}$ tend to be found at higher redshifts ($\langle z \rangle \approx 0.35$) where galaxies have molecular gas fractions ($f_{\rm H_2} \equiv M_{\rm H_2} / M_\star$) about twice higher than the local ones \citep{Tacconi2018},meaning that the possible increase of $L'_{\rm CO(2-1)}$ with $\lambda L_{5100}$ could be due to the cosmic evolution of the molecular gas content in galaxies.\footnote{There is no significant bias with host galaxy redshift when considering the Eddington ratio estimates.} To test this possibility,  we correct the host galaxy CO(2--1) luminosity by adopting the parametrization of \citet{Tacconi2018} for the cosmic evolution of $f_{\rm H_2}$ for main-sequence galaxies and a constant $\alpha_{\rm CO}$ value.  In this case,  we find no correlation between the corrected $L'_{\rm CO(2-1)}$ and $\lambda L_{5100}$ ($\tau = 0.11$,  $p=0.06$).

\subsection{The $L_{\rm IR} - L'_{\rm CO(2-1)}$ relation}
\label{sec:LIR_LCO}

\begin{figure}
\centering
\includegraphics[width=1.0\columnwidth]{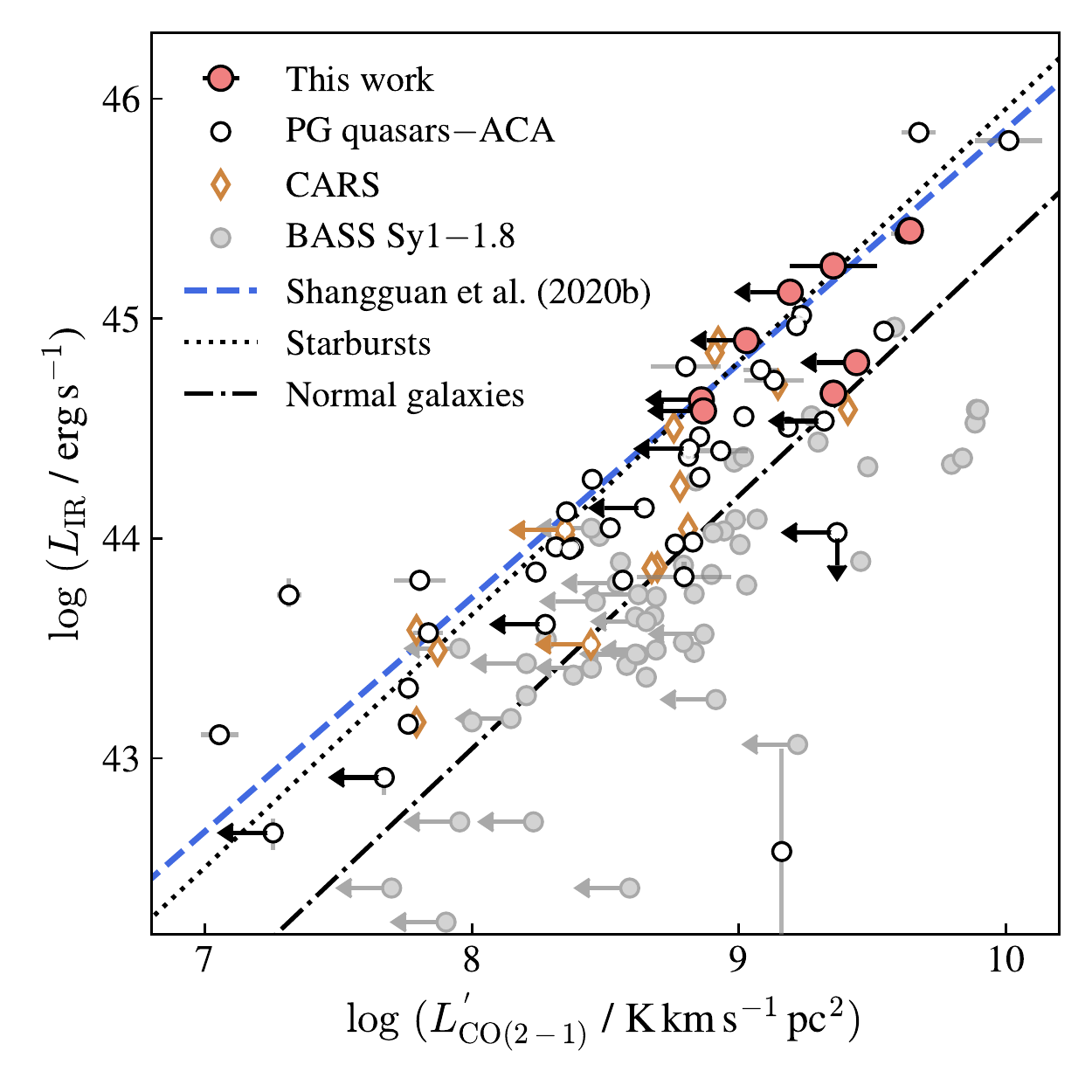}
\caption{\label{fig:LIR_LCO} Infrared luminosity as a function of the CO(2--1) line luminosity. The arrows denote $3\sigma$ upper limits. The best-fit of \citet{Shangguan2020} has a slope consistent with unity and an intrinsic scatter of $\sim0.3\,$dex. The trends for inactive systems correspond to those reported in \citet{Genzel2010} for normal star-forming systems (0.33\,dex scatter) and starbursts (0.21\,dex scatter). }
\end{figure}

In Figure~\ref{fig:LIR_LCO} we show the relation between $L_{\rm IR}$ and $L'_{\rm CO(2-1)}$ for the AGNs in our sample. We find that the more powerful quasars observed in this work ($\lambda L_{5100} \gtrsim 10^{45.5}\,$\,erg\,s$^{-1}$) follow the best-fit relation of \citet[their Equation 4, after correcting by $R_{21}$]{Shangguan2020b}. The five systems with $L'_{\rm CO(2-1)}$ upper limits provided by NOEMA tend to be on or above the best-fit relation of \citet{Shangguan2020b}, suggesting even lower $L'_{\rm CO(2-1)}$ values. However, we caution that due to the presence of strong synchrotron emission and Herschel photometry upper limits, their $L_{\rm IR}$ estimates are somewhat more uncertain than for the rest of our targets. 

\begin{figure*}
\centering
\includegraphics[width=2.0\columnwidth]{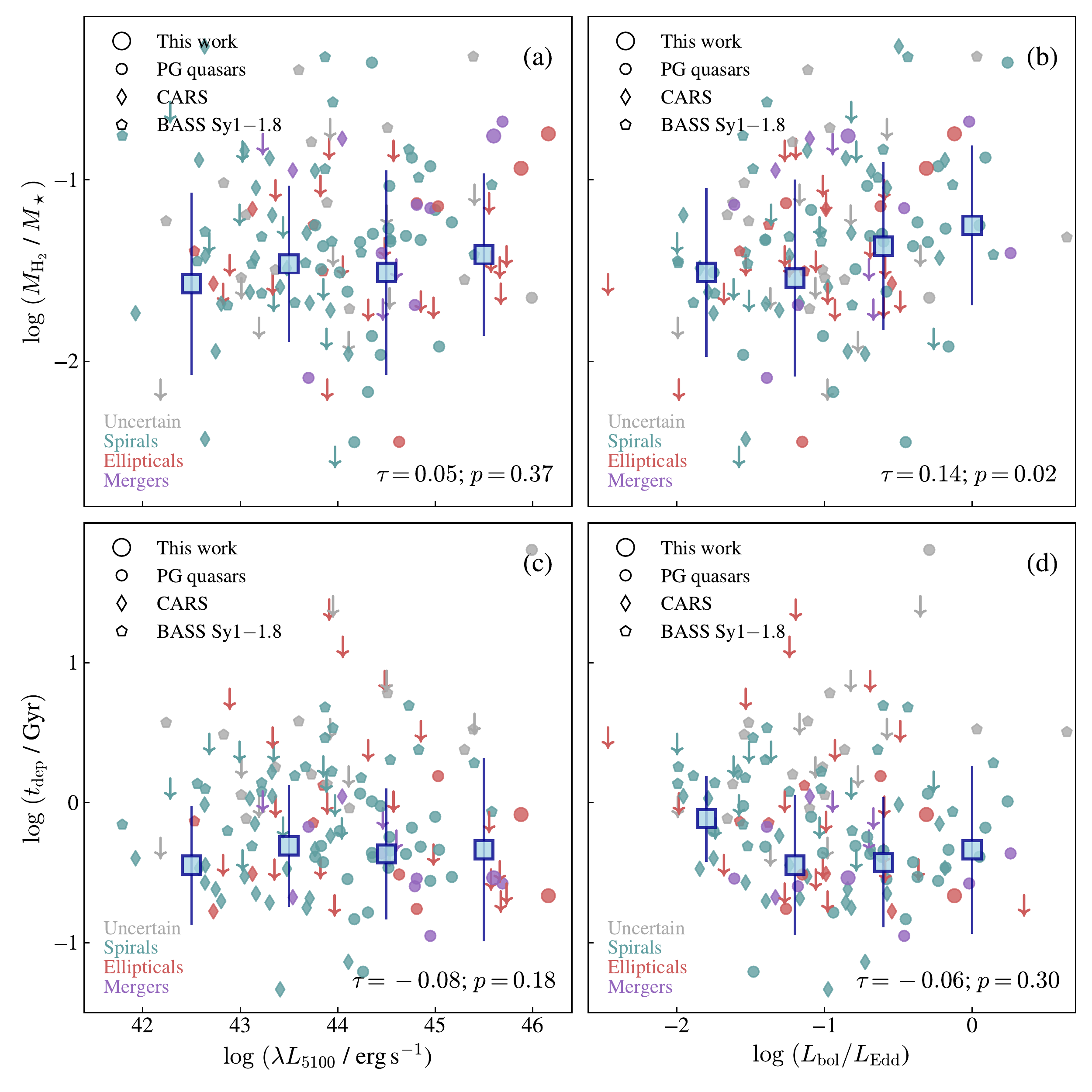}
\caption{\label{fig:fgas_tdep_AGNprops} Host galaxy molecular gas properties as a function of AGN properties.  The molecular gas fraction ($M_{\rm H_2}\,/\, M_\star$) dependence with (a) $\lambda L_{5100}$ and (b) Eddington ratio.  Panels (c) and (d) show similar dependence on AGN properties, but for the molecular gas depletion time ($M_{\rm H_2}\,/\,$SFR). In all panels, the data are colored by host galaxy morphology, and the blue squares correspond to median $M_{\rm H_2}\,/\, M_\star$ or $t_{\rm dep}$ values computed in $0.6\,$dex $\lambda L_{5100}$ and $0.3\,$dex $L_{\rm bol}/L_{\rm Edd}$ bins, respectively. The arrows correspond to $3\sigma$ upper limits.}
\end{figure*}

Surprisingly, we find that the BASS sample of type~1 AGNs is in offset with respect to the $L_{\rm IR}-L'_{\rm CO(2-1)}$ best-fit provided by \citet{Shangguan2020b} for the PG quasars.  We quantify this offset $= -0.54\,$dex in terms of $L_{\rm IR}$.  We report a similar result when including the BASS type~2 AGN data (not shown). The CARS host galaxies follow a trend similar to that reported for the PG quasars, but this result could be the affected by IR luminosity bias induced when discarding the systems with poorly sampled far-IR SED.  The redshift difference between the BASS AGN and PG quasar samples is not high enough to explain this offset.  Among the PG quasar sample, 13 out of 48 host galaxies are at $z \sim 0.05$, similarly as the BASS AGN systems, while 27 PG quasar host galaxies are at $z \lesssim 0.1$, making these practically immune to any cosmic evolutionary effect. We also note that the offset for the BASS type~1 AGNs cannot be explained by incorrect aperture correction when deriving the CO(2--1) luminosity \citep{Koss2021}. The NOEMA- and ACA-based CO(2--1) emission-line measurements for the PG quasars do not suffer from any systematics in aperture \citep{Shangguan2020b}.  Nor does the offset arise from the use of different far-IR SED models when estimating $L_{\rm IR}$.  The PG quasar far-IR modeling is detailed in Section~\ref{sec:SED_modeling}, and here we only highlight that this method is consistent with fitting a modified black-body (MBB) to the far-IR SED (see Appendix of \citealt{Shangguan2018}).  For the BASS sample, the far-IR luminosity was computed mainly by \citet{Shimizu2017}, with some few systems taken from \citet{Ichikawa2019}. Both methods are consistent between each other (see Appendix of \citealt{Shimizu2017}).  We fitted the far-IR SED of our targets following the method outlined in \citet{Shimizu2017},  namely by considering a MBB plus an additional sub-component modeling the emission from the warm dust heated by the AGN and intense star formation activity.  We found no significant discrepancy between the $L_{\rm IR}$ values provided by those fits and the estimates reported in Table~\ref{tab:CO_prop},  notwithstanding a scatter of $0.2\,$dex.  For the CARS sample, we obtain similar results when comparing the IR luminosity reported by \citet{Smirnova-Pinchukova2022} and the result of our MBB fits.  Therefore,  we find that the BASS AGNs tend to be offset with respect to the PG sample in the $L_{\rm IR}-L'_{\rm CO(2-1)}$ plane.

Compared to the non-active systems, represented by the fits for normal star-forming galaxies and starbursts \citep{Genzel2010},  the less-luminous PG quasars follow the relation of starbursts \citep{Shangguan2020b,Shangguan2020}.  The same picture holds for the most luminous PG quasars at $z \lesssim 0.5$.  The sub-sample of BASS type~1 AGNs follows the relation of normal star-forming galaxies.

\subsection{Molecular Gas Fraction and Depletion time} 
\label{sec:fgas_tdep_AGNprops}

Figure~\ref{fig:fgas_tdep_AGNprops} shows the molecular gas fraction as a function of $\lambda L_{5100}$ and Eddington ratio.  We find no significant correlation between $f_{\rm H_2}$ and $\lambda L_{5100}$ ($\tau = 0.05$,  $p$-value = 0.37).  This result does not change after taking into account the host galaxy redshift bias with $\lambda L_{5100}$ ($\tau = -0.07$,  $p$-value = 0.21). The constant $f_{\rm H_2}$--$\lambda L_{5100}$ median trend further highlights this.  On the other hand,  we find a weak correlation between $f_{\rm H_2}$ with Eddington ratio ($\tau = 0.14$,  $p$-value = 0.02),  but the corresponding $f_{\rm H_2}$--Eddington ratio median trend does not suggest any significant correlated increase.
 
Figure~\ref{fig:fgas_tdep_AGNprops} also presents the depletion time ($t_{\rm dep} \equiv M_{\rm H_2} / $SFR) as a function of $\lambda L_{5100}$ and Eddington ratio.  We find no correlations with $\lambda L_{5100}$ ($\tau = -0.08$,  $p$-value = 0.18) or Eddington ratio ($\tau = -0.06$,  $p$-value = 0.30).  Considering the depletion time cosmic evolution \citep{Tacconi2018} has a negligible effect on the correlation coefficient values.  If anything, the $t_{\rm dep}-L_{\rm bol}/L_{\rm Edd}$ median trend suggests a somewhat minor $t_{\rm dep}$ decrease from $\sim$1.0\,Gyr to $\sim$0.4\,Gyr at $L_{\rm bol}/L_{\rm Edd} \gtrsim 0.1$,  but note that the median trend may only reflect the Eddington ratio range at which the CARS and PG quasar sample dominate the median bin statistics. 

\section{Discussion}
\label{sec:dis}

The properties of the molecular gas in AGN host galaxies provide important clues regarding the role of BH accretion in shaping the evolution of galaxies. Early studies already suggested that the local luminous quasars are hosted in galaxies with significant gas reservoirs \citep{Scoville2003,Evans2006,Bertram2007}.  These findings were subsequently supported by the many new observations detecting CO emission for broader samples of local type~1 and type~2 AGNs \citep{Husemann2017,Rosario2018,Shangguan2020,Koss2021,RamosAlmeida2022,Salvestrini2022}.  The large amounts of cold gas content detected in type~1 AGN hosts are at odds with the idea of negative AGN feedback affecting the host galaxies globally (e.g., \citealt{Husemann2017,Jarvis2020,Shangguan2020b},  but see \citealt{Ellison2021,Ward2022}), with conflicting results regarding how the AGN properties correlate with the molecular gas properties in the AGN host galaxies.  \citet{Koss2021} reported that the BASS AGNs with higher Eddington ratio tend to have on average larger molecular gas reservoir,  which disagrees with the findings of \citet{Shangguan2020b} who report that the CO luminosity does not correlate with Eddington ratio for the local PG quasars (see also \citealt{Husemann2017}).  This disagreement seems to arise because both samples are complementary in terms of AGN luminosity,  with the PG quasars representative of more luminous AGNs (Figure~\ref{fig:comp_sample}).  It should be noted that \citet{Shimizu2016} already suggested that the BASS AGN sample may not extend to high enough AGN luminosity to reveal correlations with the host galaxy properties.  Conversely,  artificial correlations between host galaxy and AGN properties could also arise because of sample selection bias, with targets covering a narrow range in terms of AGN properties \citep{Zhuang2020}. When combining the BASS type~1 AGNs,  CARS, and the PG quasar samples,  including our observations,  we further confirm the lack of correlation between CO luminosity (i.e., molecular gas mass) with Eddington ratio, in agreement with \citet{Husemann2017} and \citet{Shangguan2020b}.  On the other hand, we find a weaker correlation between CO emission and AGN luminosity than that reported by \citet{Shangguan2020b},  mainly because the inclusion of the less luminous BASS and CARS AGN data to the analysis.  This correlation is mainly driven by the more luminous AGNs ($\lambda L_{5100} \gtrsim 10^{45} \,$erg\,s$^{-1}$) observed at higher redshift,  and it vanishes when correcting the CO luminosity by the cosmic evolution of the molecular gas content in galaxies.  The observed trends suggest the lack of negative AGN feedback effects on the molecular gas content of the host galaxies.

Even though we report a lack of correlation between $L'_{\rm CO(2-1)}$ and $\lambda L_{5100}$ for the full sample of type~1 AGNs (after correcting by redshift),  we note that there is still a stronger link remaining for the PG quasar sample alone \citep{Shangguan2020b}. Such trend is reminiscent of that found between AGN luminosity and SFR for local AGNs, where the processes that drive and regulate most of the star formation in low-luminosity AGN hosts presumably do not strongly affect the BH fueling, while in high luminosity AGNs it may be the opposite \citep{Rosario2012}.  If there is any correspondence between the global SFR and BH accretion events in high luminosity AGNs,  it should occur over timescales short compared to the AGN duty cycle ($\sim10-100$\,Myr; \citealt{Hopkins2005}),  requiring that most of the star formation activity being triggered on the central region of the host galaxies.  Under this scenario, the $L'_{\rm CO(2-1)}-\lambda L_{5100}$ trends reported for the full sample of type~1 AGNs and PG quasar sample alone may reflect how differently the gas is distributed within the PG quasar hosts compared with the BASS AGN host galaxies,  in addition to a closer connection between the AGN properties and the molecular gas content on the central scales of the quasar hosts.  Note that the more gas-rich PG quasars have compact molecular gas distributions \citep{Molina2021}.  In AGN hosts, the cold dust emission mainly arises from compact nuclear scales \citep{Mushotzky2014,Lutz2018,Molina2023}.  Additionally, our $L_{\rm IR}/L'_{\rm CO}$ measurements for the BASS AGNs and PG quasars presume a difference in the gas distribution between both type~1 AGN samples.  Similarly to \citet{Bertram2007}, we find that type~1 AGNs can be roughly separated into two populations in terms of $L_{\rm IR}/L'_{\rm CO}$ (proxy of the star formation efficiency or $t_{\rm dep}$),  with one population, most of the BASS AGNs, following the typical $L_{\rm IR}/L'_{\rm CO}$ ratio found in normal star-forming galaxies, and a second population, the PG quasars, and some CARS AGNs,  presenting $L_{\rm IR}/L'_{\rm CO}$ values comparable to those measured for luminous infrared galaxies (Figure~\ref{fig:LIR_LCO}).  It is well known that luminous infrared galaxies have molecular gas distributions mainly concentrated in their central kpc and sub-kpc regions (e.g., \citealt{Downes1998, Iono2009,Bellocchi2022}), while normal star-forming galaxies have more extended gas distributions (e.g., \citealt{Bolatto2017}).  Accounting for the difference in gas distribution for both type~1 AGN samples would imply adopting different $r_{21}$ and CO-to-H$_2$ conversion factor values when estimating the host galaxy molecular gas masses,  further decreasing the $M_{\rm H_2}$ values for the PG quasars, and amplifying both AGN sample differences in terms of $t_{\rm dep}$ or star formation efficiency.

Recent findings suggest the more luminous and more efficiently growing BHs tend to be hosted in starbursts \citep{Bernhard2016,Shangguan2020b,Xie2021}.  The position of an AGN host relative to the main-sequence of star-forming galaxies seems to be correlated with Eddington ratio (e.g., \citealt{Shimizu2016, Ellison2016,Woo2020,Zhuang2020,Torbaniuk2021,Zhuang2022}).  How does BH accretion connect to the galaxy's star formation activity on much larger scales? A straightforward possibility is the common dependence of BH growth and star formation on the molecular gas supply \citep{Jarvis2020,Shangguan2020},  replenished by the gravitational instabilities that drive the gas inward \citep{Kormendy2004,Romeo2016}.  Another possibility could be triggering of star formation due to positive AGN feedback, with in-situ star formation in AGN-driven outflows \citep{Maiolino2017,Gallagher2019}, and more luminous AGNs producing more frequent and stronger outflows (\citealt{Fiore2017,Rakshit2018,Fluetsch2019}; but see \citealt{Baron2019,Davies2020,Shangguan2020b,Molina2022,RamosAlmeida2022}).  The role of gas-rich major mergers cannot be overlooked, as they offer a natural explanation for the triggering of an AGN (e.g., \citealt{Hopkins2008}), specially for more luminous active galaxies \citep{Treister2012}. However, not all the galaxies hosting a luminous AGN present evidence of merger activity (e.g., \citealt{Koss2011,Kim2017,Zhao2019, Zhao2021}). The relative fraction of early-type and late-type low-$z$ quasar hosts depends only on stellar mass,  and not on bolometric luminosity or Eddington ratio \citep{Zhuang2022}. We highlight this in Figure~\ref{fig:fgas_tdep_AGNprops} where, apart from the mergers likely presenting shorter depletion times,  we find no trend between $t_{\rm dep}$ and the AGN properties with galaxy morphology.  The only clue that remains is the weak correlation between the Eddington ratio and the host galaxy molecular gas fraction, possibly implying an underlying role of gravitational instabilities or the fast inflow of cold gas in producing nuclear starbursts and the fueling of the BHs.

\section{Conclusions}
\label{sec:con}

We present new NOEMA observations of the CO(2--1) line for the most luminous eight Palomar-Green quasar host galaxies at $z \lesssim 0.5$. We detect CO(2--1) emission in three objects,  doubling the number of most luminous PG quasar host galaxy with line emission detection.  Combined with already published observations for active galaxies, we assemble CO measurements for a broad sample of $z \lesssim 0.5$ type~1 AGNs covering $\sim 3$ decades in AGN luminosity. This sample is used to investigate the relations between AGN properties, molecular gas properties, and star formation efficiency of the host galaxies.  The systems with non-detected CO emission are treated as censored data.  We summarize our conclusions as follows:
\\
\begin{itemize}

\item The luminous AGNs ($\lambda L_{5100} \gtrsim 10^{45.5}$\,erg\,s$^{-1}$) observed by NOEMA possesses CO emission consistent with the expectations from the far-IR emission.  Overall,  the total gas mass predicted from the cold dust emission,  and after subtracting the extended atomic gas content,  is well-matched with the molecular gas mass within an accuracy of $\sim 0.40\,$dex, with some small variation depending on the host galaxy morphological type.
\\
\item The PG quasar and CARS AGN host galaxies follow a tight,  linear $L_{\rm IR}-L'_{\rm CO}$ correlation consistent with that found for starbursts galaxies \citep{Shangguan2020}.  The BASS AGNs follow a similar correlation, but they show a zero-point offset placing them on the relation for normal star-forming galaxies. 
\\
\item We find a weak correlation between CO(2--1) line luminosity and AGN luminosity,  as well as between molecular gas fraction and Eddington ratio.  The former correlation vanishes when considering the cosmic evolution of the gas content in galaxies and that the more luminous AGN hosts are found at higher redshifts, meaning that the AGN luminosity is not correlated with host galaxy global molecular gas content.  The second correlation points to the role of host galaxy gravitational instabilities or the fast inflow of cold gas in triggering AGN activity.  Both findings suggest an ineffective role of negative AGN feedback.
\end{itemize}
\
\acknowledgments{We thank the anonymous referee for helpful comments and suggestions.  We acknowledge support from the National Science Foundation of China (11721303, 11991052, 12011540375, 12233001) and the China Manned Space Project (CMS-CSST-2021-A04, CMS-CSST-2021-A06).  R.W. acknowledges support from the National Science Foundation of China grants No. 12173002. This work was funded by ANID - Millennium Science Initiative Program - ICN12\_009 (FEB), CATA-BASAL -  FB210003 (FEB, ET), and FONDECYT Regular - 1190818 (ET, FEB) and 1200495 (FEB, ET).  We thank to Stefano Antonellini for his support during the NOEMA data reduction process.  This work is based on observations carried out under project number W21CI with the IRAM NOEMA Interferometer. IRAM is supported by INSU/CNRS (France), MPG (Germany) and IGN (Spain)}

\software{\textsc{Astropy}\,\citep{astropy:2013,astropy:2018}, \textsc{matplotlib}\,\citep{Hunter2007}, \textsc{numpy}\,\citep{Oliphant2006}, GILDAS\,\citep{Guilloteau2000}.}
\\
\\

\appendix

\section{Radio Continuum Measurements}
\label{sec:AppA}

Radio continuum measurements collated from the literature for the eight sources presented in this work. 

\begin{table*}
        \centering
        \def\arraystretch{1.0}
        \setlength\tabcolsep{3pt}
        \caption{\label{tab:radio_data} Radio Continuum Measurements}
        \vspace{0.2mm}
        \begin{tabular}{cccc}
                \hline
                \hline
                Object & Frequency & $S_{\nu}$ & Reference \\
                & (GHz) & (mJy) & \\
                (1) & (2) & (3) & (4) \\
                \hline
        PG\,1049$-$005 & 1.4 & $0.82\pm0.14$ & \citet{Becker1995}  \\
        &5.0 & $0.48\pm0.03$ & \citet{Kellermann1989}  \\
	   & 8.5 & $0.37\pm0.07$ & \citet{Baldi2022} \\
        & 45 & $< 0.18^*$ & \citet{Baldi2022} \\
	   PG\,1100+772 & 1.4  & $2341\pm79$ & \citet{Condon1998} \\
	   & 2.7  & $1415\pm40$ & \citet{Laing1980} \\
	   & 5.0  & $775\pm30$ &  \citet{Laing1980}\\
	   & 10.7 & $375\pm19 $ & \citet{Laing1980}\\
	   & 14.9 & $310\pm20$ &  \citet{Laing1980} \\
	   & 31.4 & $330\pm130$ & \citet{Geldzahler1983} \\
	   PG\,1259+593 & 1.4 & $< 0.45$ &  \citet{Condon1998} \\
	   & 5.0 & $< 0.3^*$ &  \citet{Baldi2022} \\
	   & 45 & $< 0.15^*$ & \citet{Baldi2022} \\
	   PG\,1425+267 &	0.074 &	 $2710\pm360$ & \citet{Cohen2007} \\ 
	   & 0.151 & $2260\pm158$ & \citet{Waldram1996} \\  
	   & 0.365 & $572\pm64$ & \citet{Douglas1996}  \\  
	   & 1.4 & $108\pm4$ & \citet{Condon1998} \\ 
	   & 4.85 & $112\pm16$ &  \citet{Gregory1991} \\  
	   PG\,1512+370	 &	0.074 & $9860\pm1010$ & \citet{Cohen2007} \\ 
	   & 0.365 & $3003\pm49$ & \citet{Douglas1996} \\ 
	   & 1.4 & $990\pm31$ & \citet{Condon1998} \\
	   & 4.85 & $361\pm45$ &  \citet{Gregory1991} \\
	   PG\,1704+608 & 1.49 & $1258\pm123$ & \citet{Kuehr1981} \\ 
	   & 4.85 & $590\pm30$ & \citet{Gregory1991} \\ 
	   & 10.7 & $420\pm20$ & \citet{Kuehr1981} \\
	   PG\,2112+059	 & 1.4 & $3.0\pm0.4$ & \citet{Condon1998} \\
	   & 1.49 & $2.93\pm0.38$ & \citet{Barvainis1996}	 \\
	   & 4.89 & $0.89\pm0.09$ & \citet{Barvainis1996} \\
	   & 4.9 & $0.89\pm0.07$ &	\citet{Kellermann1989} \\
	   & 8.4 & $0.52\pm0.05$ & \citet{Barvainis2005} \\	   
	   & 8.48 & $0.56\pm0.07$ & \citet{Barvainis1996} \\
	   & 14.9 & $< 0.48$ & \citet{Barvainis1996} \\
	   PG\,2251+113	 &	0.074 & $11840\pm1240$ & \citet{Cohen2007} \\ 
	   & 0.365 & $4387\pm89$ & \citet{Douglas1996} \\ 
	   & 1.4 & $1436\pm43$ & \citet{Condon1998} \\  
	   & 2.7 & $700$ & \citet{Wright1990} \\ 
	   & 4.85 & $615\pm85$ & \citet{Gregory1991} \\
        \hline
        \end{tabular}
        \justify
        {\justify \textsc{Note}--- (1) Source name. (2) Restframe frequency. (3) Flux density. (4) Reference.  If available, we provide a $3\sigma$ flux density upper limit. ($^*$) Upper limit values were calculated by using the observation rms as input to Equation~\ref{eq:uplim_cont} and assuming an aperture size equal to 3 times the observation synthesized beam size ($N_{\rm A}/N_{\rm B}=3$).}
\end{table*}

\bibliography{bibliography}
\bibliographystyle{aasjournal}
\end{document}